\documentclass[10pt,a4paper]{article}
\pdfoutput=1
\usepackage[english]{babel}
\usepackage[latin1]{inputenc}
\usepackage{amsfonts,amsbsy,bm,euscript,mathrsfs}
\usepackage{amssymb,stmaryrd,faktor,slashed}
\usepackage{color}
\usepackage[tbtags]{amsmath}
\usepackage[bookmarks=true,colorlinks=true,linkcolor=black,citecolor=black,urlcolor=black,bookmarksnumbered]{hyperref}
\usepackage[nosort]{cite}

\numberwithin{equation}{section}

\makeatletter
\renewcommand\section{\@startsection {section}{1}{\z@}
{-3.5ex \@plus -1ex \@minus -.2ex}
{2.3ex \@plus.2ex}
{\normalfont\Large\bfseries}}
\renewcommand\subsection{\@startsection{subsection}{2}{\z@}
{-3.25ex\@plus -1ex \@minus -.2ex}
{1.5ex \@plus.2ex}
{\normalfont\large\bfseries}}
\makeatother

\newcommand{\arxivlink}[1]{\href{http://arxiv.org/abs/#1}{\tt arXiv:#1}}
\newcommand \foot [1] {\footnote{#1\vspace{2pt}}}
\newcommand \rf [1] {(\ref{#1})}

\def \texpdf {\texorpdfstring}
\def \be {\begin{eqnarray}}
\def \ee {\end{eqnarray}}

\def \bi{\bibitem}

\def \td {\tilde}
\def \ci{\cite}

\def \z {\zeta}

\def \a {\alpha}
\def \b {\beta}
\def \e {\varepsilon}

\def \del {\partial}
\def \a {\alpha}

\def \z {\zeta}

\def \ov {\over}

\def \b {\beta}
\def \l {\lambda}

\def \ci {\cite}

\def \P {\Phi}
\def \l {\lambda}

\def \td {\tilde}

\def \m {\mu}

\def \e {\epsilon}
\def \bi{\bibitem}
\def \la {\label}
\def \l {\lambda}

\def \adss {$AdS_5 \times S^5~$ }
\def \ov {\over}

\def \r {\rho}
\def \no {\nonumber}

\def \del {\partial}

\def \bi {\bibitem}
\def \la {\label}
\def \l {\lambda}

\def \adss {$AdS_5 \times S^5$\ }

\def \r {\rho}

\def \ov {\over}

\def \varpi {{\rm w}}

\def \n {\nu}

\def \vp {\varphi}

\def \CC {{\rm C}}

\def \eea {\ee}
\def \eqref {\rf}

\def \adss {$AdS_5\times S^5$\ }
\def \adt {$AdS_2 \times S^2$\ }
\def \iffa {\iffalse}
\def \ba {\be}

\def \adt {$AdS_2 \times S^2$\ }
\def \adst {$AdS_3 \times S^3$\ }

\def \emo {$\eta$-model\ }
\def \lmo {$\l$-model\ }

\def \vk {\varkappa}

\def \xxp {x^-}
\def \xxm {x^+}
\def \np {\newpage}
\def \W {\land}
\def \ot { {\hat t}}
\def \ovarphi {{\hat \varphi}}
\def \opsi {{\hat \psi}}
\def \ophi {{\hat \phi}}

\def \reo {\color{black}}

\def \adt {$AdS_2 \times S^2$\ }
\def \adst {$AdS_3 \times S^3$\ }

\def \CC {{\cal C}}
\def \X {{\rm F}}
\def \tpsi {{\tilde \psi}}
\def \eE {{\bf \rm e}}

\begin{document}

\setcounter{equation}{0}
\setcounter{footnote}{0}
\setcounter{section}{0}

\vspace{-3cm}
\thispagestyle{empty}
\vspace{-1cm}

\rightline{ HU-EP-15/34}
\rightline{ Imperial-TP-AT-2015-05}

\begin{center}
\vspace{1.5cm}

{\Large\bf Type IIB supergravity solution for the T-dual \\
of the $\eta$-deformed \adss superstring \\ \vspace{0.2cm}}

\vspace{1.5cm}

{B. Hoare$^{a,}$\footnote{ben.hoare@physik.hu-berlin.de} and A.A. Tseytlin$^{b,}$\footnote{Also at Lebedev Institute, Moscow. tseytlin@imperial.ac.uk }}\\

\vskip 0.6cm

{\em $^{a}$ Institut f\"{u}r Physik und IRIS Adlershof, Humboldt-Universit\"at zu Berlin,
\\ Zum Gro\ss en Windkanal 6, 12489, Berlin, Germany.}

\vskip 0.3cm

{\em $^{b}$ The Blackett Laboratory, Imperial College, London SW7 2AZ, U.K.}

\vspace{.2cm}
\end{center}

\begin{abstract}
We find an exact type IIB supergravity solution that represents a one-parameter
deformation of the T-dual of the $AdS_5 \times S^5$ background (with T-duality
applied in all 6 abelian bosonic isometric directions). The non-trivial fields
are the metric, dilaton and RR 5-form only. The latter has remarkably simple
``undeformed'' form when written in terms of a ``deformation-rotated'' vielbein
basis. An unusual feature of this solution is that the dilaton contains a
linear dependence on the isometric coordinates of the metric precluding a
straightforward reversal of T-duality. If we still formally dualize back, we
find exactly the metric, $B$-field and product of dilaton with RR field
strengths as recently extracted from the $\eta$-deformed $AdS_5 \times S^5$
superstring action in arXiv:1507.04239. 
We also discuss similar solutions for deformed $AdS_n \times S^n$ backgrounds
with $n=2,3$. In the $\eta \to i$ limit we demonstrate that all these
backgrounds can be interpreted as special limits of gauged WZW models and are
also related to (a limit of) the Pohlmeyer-reduced models of the $AdS_n \times
S^n$ superstrings.
\end{abstract}

\newpage
\setcounter{equation}{0}
\setcounter{footnote}{0}
\setcounter{section}{0}

\tableofcontents

\section{Introduction}\label{secint}

Finding the supergravity background corresponding to the $\eta$-deformation of \adss superstring sigma model 
\ci{dmv,klim} (``$\eta$-model'') has turned out to be a non-trivial problem. The corresponding metric and $B$-field
were read off the superstring action in \ci{abf}. It was then found that for low-dimensional analogs \ci{hrt} of the deformed metric, corresponding to deformations of the \adt and \adst
sigma models, it is possible \ci{lrt} to find special combinations of RR fluxes and the dilaton that complete the metrics to
full type IIB supergravity solutions. However, it was noticed that there may be many dilaton/flux backgrounds supporting the same metric and it was not checked that the solutions that were found (with particular non-factorized dilatons) correspond to the $\eta$-deformed \adt and \adst superstring sigma models. Very recently, the RR background that follows directly from
quadratic fermionic term in the $\eta$-deformed \adss sigma model was finally found in \ci{abf2}
but surprisingly it was
found
that the resulting metric, $B$-field and RR fluxes
cannot be supported by a dilaton to promote them to a consistent type IIB supergravity solution.

In a parallel development,
a one-parameter deformation of the non-abelian dual of the \adss superstring sigma model was constructed \ci{hms}
(generalizing the bosonic
models of
\ci{sfet}). This ``$\l$-model'' is closely connected
(via an analytic continuation)
to the $\eta$-model at the classical phase space level (the associated Poisson bracket algebras are effectively isomorphic \ci{dmv2}).
Furthermore, it was found in \ci{ht} that the metric of the $\eta$-model
can be obtained from the metric of the $\l$-model by a certain singular limit
(involving infinite shifts of coordinates corresponding to Cartan directions of the original symmetry group) and an analytic continuation relating the two deformation parameters ($\eta = i {1-\l\ov 1+\l}$).

More precisely, the metric that originated from this singular limit of the $\l$-model metric was not the $\eta$-deformed \adss metric itself but its T-dual with respect to all 6 isometric directions associated to Cartan generators of $SO(2,4) \times SO(6)$. The reason for this can be traced to the fact that the $\l$-deformation was applied to the non-abelian T-dual of the \adss coset model and performing the non-abelian duality
implies dualizing with respect to the whole symmetry group.
Applying the singular limit
gives preference to the Cartan directions, such that it
should produce a
deformation of the abelian T-dual of the \adss model \ci{ht}.
This observation \ci{ht} of the
special role
of the T-dual of the deformed \adss model turns out to be crucial in what follows.

Guided by the existence of a supergravity solution that supports
the metric of the \adss
$\l$-model by a particularly simple (factorized) dilaton and just
the RR 5-form flux \ci{Sfetsos:2014cea,dst}, in \ci{ht} we
applied the above singular limit to its \adt counterpart and found
a new supergravity solution (different to the T-dual of the solution found
in \ci{lrt})
that supports the T-dual of the $\eta$-deformed \adt metric by a
RR 2-form flux and factorized dilaton.
A peculiar feature of this solution was that the dilaton
contained a term
linear in the two isometric coordinates of the metric. This precluded us from applying the standard rules to reverse the T-duality and
find the supergravity background supporting the original $\eta$-deformation of \adt metric.

Motivated by the observations of \ci{ht}, here we directly construct similar type IIB supergravity
solutions supporting the T-duals
of the $\eta$-deformations of the \adst and \adss metrics.
Again, the resulting dilatons contain terms linear in (some of) the
isometric coordinates, which precludes us from undoing
the T-duality and thus finding similar solutions
supporting the $\eta$-deformed \adst and \adss metrics themselves.

The solution we find for the T-dual of the $\eta$-deformed \adss metric
contains only the dilaton $\Phi$ and the RR 5-form flux $F_5$.
{Surprisingly, if we formally apply the standard T-duality rules \ci{hull,hass,tds}
to this background, we obtain the metric, $B$-field {\it and} precisely the
product of the mixed RR fluxes with the dilaton, $e^\Phi F_n$,
as found directly from the $\eta$-deformed \adss sigma model action in \ci{abf,abf2}.} Since this T-dualization can be done explicitly at the level of the classical string action (ignoring the issue of the quantum dilaton shift)
this supports the idea that the solution we find
is the one associated with the $\eta$-model.

This also explains the conclusion of \ci{abf2} that the
background extracted from the $\eta$-model action
cannot be promoted to a supergravity solution for any choice of the dilaton. Indeed, the usual expectation
that T-duality should map from one supergravity solution to another does not apply
in cases in which the dilaton explicitly
depends on the isometries of the metric -- the new T-dual dilaton will depend on the original isometric coordinates, while the dual metric and other fields will
describe their dual analogs. 
One might attempt to interpret the resulting background
as a solution of some ``doubled'' version of type IIB string theory
where both the original and dual coordinates are treated on an equal footing \ci{ddd},
or, possibly, of ``doubled'' type IIB supergravity \ci{doub} but with the strong constraint relaxed.\foot{The
usual discussions of ``doubled'' field theory assume the weak $\del_i {\td \del}^i X =0$
as well as the strong constraint $\del_i X {\td \del}^i Y =0$
for any two fields $X,Y$ (cf., however, \ci{nun}). The former is satisfied in our case while the latter is
not as we have ${\td \del}^i \Phi \not=0$ for the dilaton or RR flux while $\del_i g \not=0$ for the metric.}
An alternative is to search for a T-dual solution where the dilaton depends
linearly on the same dual coordinates that appear in the metric, i.e. to map
the ``momentum"   mode of the dilaton into the ``winding" one.  We leave an
investigation of these ideas for the future.

To sum up, here we will show that,
while it is presently still unclear how to directly interpret the background found \ci{abf,abf2} from the \adss $\eta$-model
as a type IIB supergravity solution, the background formally related to it
by T-duality in all 6 isometric directions can indeed be promoted to an exact supergravity solution by properly adjusting the dilaton
(in particular, adding terms linear in some isometric coordinates).
We shall also provide another interpretation of these linear dilaton terms in the special $\eta \to i$
limit by showing that they appear from the dilaton of the standard gauged WZW model upon taking a special limit required to obtain the $\eta$-deformed metric as in \ci{ht}.

\

We shall start in section \ref{secsugra2} with a review of the
solution of type IIB supergravity
for the T-dual of the $\eta$-deformed \adt metric supported by a
factorized, non-isometric dilaton and just a single imaginary RR 2-form flux
(originating from the 5-form in 10d supergravity upon compactification on $T^6$) \ci{ht}.
Then in sections \ref{secsugra3} and \ref{secsugra5}
we shall construct the analogous backgrounds in the \adst and \adss cases. These solutions will possess the same features, i.e. they will be supporting the T-dual
of the $\eta$-deformed metric with a
factorized, non-isometric dilaton and
just a single imaginary RR 3-form flux for \adst and self-dual 5-form flux for \adss.
{ In the
$AdS_5 \times S^5$ case we shall explicitly check (in Appendix \ref{appa})
that after formally T-dualizing
along the isometric directions of the metric
we recover the background fields (metric, $B$-field and $e^\Phi F$) of the
supercoset \emo of \cite{dmv}, which were found in \cite{abf2}. }

Furthermore, these backgrounds should appear as limits of the \lmo backgrounds constructed
in \cite{Sfetsos:2014cea} and \cite{dst}. In section \ref{3.1} we provide evidence
for this in the direct $\eta \to i$ or, equivalently, $\vk \equiv { 2 \eta \ov 1 - \eta^2} \to i$
limit, in which the RR fluxes vanish.
In section \ref{3.2} we shall consider a refined $\vk\to i$ limit in which one also
rescales the ``longitudinal'' coordinates, resulting in a
pp-wave background \ci{hrt,ht}, which
is related to
the Pohlmeyer-reduced model.
Some concluding remarks will be made in section 4.

\section{Supergravity backgrounds for T-duals to \texpdf{$\eta$}{eta}-deformed \texpdf{$AdS_n \times S^n$}{AdSn x Sn} models}\label{secsugra}

We shall consider the deformed models for $AdS_n \times S^n$ with $n=2,3, 5$ in parallel.
The undeformed \adss metric is a solution of type IIB supergravity with constant dilaton and homogeneous $F_5$ flux. Applying T-duality in all 3+3 isometric directions
we formally arrive at another supergravity solution with non-constant dilaton and
(since T-duality is applied in the time direction \ci{Hull}) an
{\it imaginary} 5-form flux. Similarly, starting with the $AdS_2 \times S^2$ ($AdS_3 \times S^3$) solution of
type II supergravity compactified on $T^6$ ($T^4$) we again find a solution
supported by a non-trivial dilaton and imaginary 2-form (3-form) flux in the effective 4d (6d) supergravity.

One may then look for solutions which represent deformations of these T-dual $AdS_n \times S^n$
backgrounds, i.e. such that their metrics are the same as T-duals of the $\eta$-deformed
$AdS_n \times S^n$ metrics in \ci{abf,hrt}.
As in \ci{abf,hrt,ht} we shall use
\be \vk \equiv { 2 \eta \ov 1 - \eta^2} \la{01} \ee
as the deformation parameter in the supergravity fields.
The minimal assumption is that such solutions
should be again supported just by a dilaton and a single (imaginary) RR $n$-form.
Indeed, we shall find such solutions but, as discussed in the Introduction,
we will be unable to dualize back to get
real backgrounds due to linear terms in the dilaton present for $\vk\not=0$, which
break the isometries of the undeformed background.

\subsection{\texpdf{$AdS_2 \times S^2$}{AdS2 x S2}}\label{secsugra2}

Let us start by reviewing the supergravity solutions for the \emo in the
$AdS_2 \times S^2$ case found in \cite{ht}, which were constructed by taking limits
of the \lmo backgrounds presented in \cite{Sfetsos:2014cea}.

Here we compactify 10d type IIB supergravity on $T^6$ to four
dimensions retaining the metric, dilaton and a single RR 2-form flux.
The field content of the corresponding truncation of the 10d supergravity is
given by the metric, dilaton and self-dual RR 5-form, which is built from the
2-form flux $F$ and the holomorphic 3-form on $T^6$, see, for example,
\ci{sor,lrt}. The resulting bosonic 4d action is then given by
\begin{equation}\label{sact2}
\mathcal{S}_2 = \int d^4x \; \sqrt{-g}\Big[e^{-2\Phi}\big[R + 4(\nabla \Phi)^2\big] - \frac14 F_{mn}F^{mn}\Big]\ .
\end{equation}
The corresponding equations of motion and Bianchi identities are ($m,n,\ldots =0,1,2,3$)
\begin{align}\nonumber
& R + 4 \nabla^2 \Phi - 4 (\nabla\Phi)^2 = 0\ ,
&& R_{mn} + 2 \nabla_m\nabla_n \Phi = \frac{e^{2\Phi}}2 (F_{mp}F_n{}^p - \frac14 g_{mn}F^2) \ ,
\\\label{seom2}
& \partial_n (\sqrt{-g}F^{mn}) = 0 \ ,
&& \partial_{[p} F_{mn]} = 0 \ .
\end{align}
The first two equations imply that the dilaton should satisfy $\nabla^2 e^{-2 \Phi} =0$.

The metric of the \emo in this case is given by \ci{abf,hrt}
\begin{equation}\label{met2}
d{\hat s}_2^2 = -\frac{1+\rho^2}{1-\varkappa^2\rho^2} d\ot^2 + \frac{d\rho^2}{(1-\varkappa^2\rho^2)(1+\rho^2)}
+\frac{1-r^2}{1+\varkappa^2r^2}d\ovarphi^2 + \frac{dr^2}{(1+\varkappa^2r^2)(1-r^2)}\ ,
\end{equation}
which has a $U(1)^2$ isometry,\foot{Here and below we shall formally refer to abelian isometries corresponding to
translations in some direction as ``$U(1)$ isometries'', i.e. we will not distinguish between
compact and non-compact isometries.}
corresponding to the rank of $PSU(1,1|2)$, given by
shifts in $\ot$ and $\ovarphi$.
We use hats to denote isometries that we will T-dualize; the corresponding
dual coordinates will have no hats.

In \cite{ht} two solutions of the equations \eqref{seom2} that support the
metric \eqref{met2}, up to T-dualities, were given.
Here we will mainly consider the solution corresponding to the metric \rf{met2}
T-dualized in both isometries
(the target-space indices $m,n,\ldots = 0,1,2,3$ correspond to $t,\r,\vp,r$)\foot{Let 
us emphasize 
that the solution of \rf{seom2}   supporting  the  T-dual to the metric \rf{met2} 
is not unique. For example, the solution of the same dilaton equation $\nabla^2 e^{-2 \Phi} =0$
corresponding to the T-dual of  the solution   found in \ci{lrt}  is different: 
$e^{\Phi_2'} = \frac{{1-\varkappa \rho\, r}  }{\sqrt{1+ \rho ^2} \sqrt{1- r^2}}$, i.e. this dilaton    is isometric but not factorizable.}
\begin{align}\nonumber
&ds_2^2 = \eta_{ab}e^a e^b \ , \qquad
e^0 = \frac{\sqrt{1-\vk^2\rho^2}}{\sqrt{1+\rho^2}} dt \ , \qquad
e^1 = \frac{d\rho}{\sqrt{1-\varkappa^2\rho^2}\sqrt{1+\rho^2}} \ ,
\\ \nonumber
& \hspace{87pt} e^2 = \frac{\sqrt{1+\vk^2r^2}}{\sqrt{1- r^2}} d\vp \ , \qquad
e^3 = \frac{dr}{\sqrt{1+\varkappa^2r^2}\sqrt{1-r^2}} \ ,
\\ & e^{\Phi_2} F_2 = \frac{\sqrt{2}i\sqrt{1+\vk^2}}{\sqrt{1-\vk^2\r^2}\sqrt{1+\vk^2r^2}}\nonumber
\big[
(e^0 \W e^3 + e^1 \W e^2)
- \varkappa^2 \rho r (e^0 \W e^3 -e^1 \W e^2 )
\\\nonumber & \hspace{175pt}
+ \varkappa \rho( e^0 \W e^2 + e^1 \W e^3)
+ \varkappa r( e^0 \W e^2 -e^1 \W e^3) \big] \ ,
\\\label{sol2}
& e^{\Phi_2} =
e^{\Phi_0 - \varkappa (t+\varphi )} \frac{\sqrt{1-\varkappa^2\rho ^2} \sqrt{1+\varkappa^2 r^2} }{\sqrt{1+ \rho ^2} \sqrt{1- r^2}}\ .
\end{align}
The RR flux actually has a remarkably simple form
that can be made explicit by introducing the
boosted/rotated zweibein bases
\be
&&\eE^t \equiv {1\ov \sqrt{1-\vk^2\r^2}}\,\big( e^0 + {\vk\r} \, e^1\big)
= {1\ov \sqrt{1+ \r^2}} \big( dt + \frac{\vk \r}{1-\vk^2\r^2} d\r\big)\ , \no\\
&&
\eE^\r \equiv {1\ov \sqrt{1-\vk^2\r^2}}\, \big( e^1 + {\vk \r} \, e^0\big)
= {1\ov \sqrt{1+ \r^2}} \big( \frac{1}{1-\vk^2\r^2} d\r + \vk \r\, dt\big)\ , \no\\
\no \\
&&\eE^\vp \equiv {1\ov \sqrt{1+\vk^2r^2}}\, \big( e^2 - {\vk r} \, e^3\big)
= {1\ov \sqrt{1- r^2}} \big( d\vp - \frac{\vk r}{1+\vk^2r^2} dr\big)\ , \no
\\
&&
\eE^r \equiv  {1\ov \sqrt{1+\vk^2r^2}}\, \big( e^3 + {\vk r} \, e^2\big)
= {1\ov \sqrt{1- r^2}} \big( \frac{1}{1+\vk^2r^2} dr + \vk r\, d\vp \big) \ , \la{n1}
\ee
such that
\begin{align}\no
ds^2_2 & = \eta_{ab} e'^a e'^b \ , \qquad \qquad
e^{\Phi_2} F_2 = {\sqrt{2}i\sqrt{1+\vk^2}}
\ \big(e'^0 \W e'^3 + e'^1 \W e'^2\big) \ ,
\\ e'^0 & = \eE^t \ , \qquad e'^1 = \eE^\r \ ,\qquad
 e'^2= \eE^\vp \ , \qquad e'^3 = \eE^r \ . \la{n2}
\end{align}
Thus the $\vk$-deformation is a ``rotation'' that preserves the structure of the undeformed
background; it only affects the definition of the tangent basis (and dilaton).
In particular, in this basis $e^{\P_2} F_2$ remains constant and is just rescaled
by a factor of $\sqrt{1+\vk^2}$.\foot{As in the undeformed limit, there is actually a one-parameter family of fluxes that
solve the supergravity equations \eqref{seom2} given by
\begin{equation*}
e^{\Phi_2} F_2 = {\sqrt{2}i\sqrt{1+\vk^2}}
\ \big(\text{c}_1\, e'^0 \W e'^3 + \text{c}_2\, e'^1 \W e'^2\big) \ , \qquad \text{c}_1^2 + \text{c}_2^2 = 2 \ .
\end{equation*}\vspace{-10pt}}
Furthermore, the RR potential  $C_1$ for  $F_2 = d C_1$
also takes a simple form in this basis\foot{Note
that the singular term in $C_1$ appearing in the limit 
$\varkappa \to 0$ is pure gauge, i.e. $C_1 \to  \sqrt{2}i \vk^{-1} ( dr - d \rho) + \ldots$. 
One can of course choose an alternative gauge in which $C_1$ is manifestly regular for $\vk \to 0$.}
\begin{equation}\la{28}
e^{\Phi_2} C_1 = \sqrt{2}i\sqrt{1+\vk^{-2}}\Big(
\frac{e'^3}{\sqrt{1+\r^2}} -
\frac{e'^1}{\sqrt{1-r^2}}
\Big) \ .
\end{equation}
Some other important features of this solution that are worth noting are:
\begin{itemize}
\item The background fields entering the classical Green-Schwarz action (the metric and $e^\Phi F_2$) are invariant
under the $U(1)^2$ isometry
given by shifts in $t$ and $\varphi$. This isometry is
broken to (a ``null'') $U(1)$ in the dilaton by the linear $t+ \vp$ term
\item For $\varkappa \in \mathbb{R}$ the metric and dilaton are real, while the RR flux is imaginary.
\item For $\varkappa \to 0$ the $U(1)^2$ isometry is restored in the full background, and we
can T-dualize back
recovering
the standard Bertotti-Robinson solution for $AdS_2 \times S^2$ (i.e. the $\varkappa\to 0$ limit of \rf{met2})
with constant dilaton and {\it real} homogeneous 2-form flux
which has a factorized ``2+2'' form.
\item For $\varkappa \to \infty$, rescaling the fields and the string tension, we find a non-standard background,
i.e. the dilaton still has a linear dependence on the isometric coordinates.
\item For $\varkappa \to i$ the flux vanishes, while the $t$ and $\varphi$ directions become free. We will discuss this limit in more detail in section \ref{kai}.
\end{itemize}

The dependence of the dilaton on $t +\varphi$ in \rf{sol2}
prohibits one directly T-dualizing in these directions to recover the metric \rf{met2}.\foot{One could still
perform T-duality in the orthogonal ``null'' direction $t -\varphi$, but that appears to give a
complicated background with an extra $B$-field, i.e. it does not bring us back to \rf{met2}.}
Still, it is interesting to note that if we formally T-dualize the metric using the standard rules
we will get an additional shift of the dilaton that will cancel the square root factors in $e^{\Phi_2}$ in \rf{sol2}.
One may thus attribute the origin of these factors to T-dualizing from \rf{met2} to \rf{sol2}.
Then we get simply
\be\la{di2}
\hat \Phi_2 = \Phi_0 - \varkappa (t + \varphi)
\ee
as the dilaton associated to \rf{met2}.
Note, however, that this dilaton depends on the dual counterparts $t$ and $\vp$ of the original coordinates
$\hat t$ and $\hat \varphi$ in \rf{met2}, i.e. the resulting background will not have an
immediate interpretation as a standard type IIB supergravity solution.

The other background considered in \cite{ht} is for the metric \eqref{met2},
i.e. with no T-dualities. This solution is related to \eqref{sol2} by the
formal map\foot{For $\varkappa^2 \in (0,-1]$, i.e. including the point $\varkappa = i$,
the map \eqref{map2} is a real diffeomorphism.}
\begin{equation}\label{map2}
t \to \frac{i \ot}{\varkappa} \ , \qquad \rho \to \frac{i}{\varkappa\rho} \ , \qquad
\varphi \to \frac{i \ovarphi}{\varkappa} \ , \qquad r \to \frac{i}{\varkappa r} \ .
\end{equation}
The corresponding dilaton and RR flux are then complex so the interpretation of this solution is unclear.
Indeed, the $\varkappa \to 0$ limit of the resulting background represents a
non-standard solution -- the undeformed $AdS_2 \times S^2$ metric supported by a complex
dilaton with a linear dependence on the isometric directions and complex RR
flux. It does, however, have a natural $\varkappa \to \infty$ limit if we first
use the rescaling
\begin{equation}\begin{split}\label{kainflim1}
& \ot \to \varkappa^{-1} \ot \ , \qquad
\rho \to \varkappa^{-1} \rho \ , \qquad
\ovarphi \to \varkappa^{-1} \ovarphi \ , \qquad
r \to \varkappa^{-1} r \ ,
\\
& ds^2 \to \vk^2 ds^2 \ , \qquad e^{\Phi_2} F_2 \to \vk e^{\Phi_2} F_2 \ , \qquad e^{\Phi_2} \to \vk^{-2} e^{\Phi_2} \ .
\end{split}\end{equation}
The resulting real background is the ``mirror'' model one
constructed in \cite{mirror0} and is related to a $dS_2 \times H^2$
solution by T-dualities in $\ot$ and $\ovarphi$,
with a constant dilaton and imaginary homogeneous RR flux.
Thus the parameter region around
$\varkappa= \infty$ corresponds
to considering the \emo for $dS_2 \times H^2$ with deformation parameter
$\widetilde \vk = \vk^{-1}$.

\subsection{\texpdf{$AdS_3 \times S^3$}{AdS3 x S3}}\label{secsugra3}

Next, let us consider the background corresponding to the \emo for $AdS_3
\times S^3$.  The solution we find can be embedded into 10d type IIB
supergravity by compactifying on $T^4$ to six dimensions retaining only the
metric, dilaton and a single RR 3-form flux.
The field content of the corresponding 10d supergravity solution
will also be given by the metric, dilaton and a RR 3-form flux.
The resulting truncated bosonic 6d action is then given by
\begin{equation}\label{sact3}
\mathcal{S}_6 = \int d^6x \; \sqrt{-g}\Big[e^{-2\Phi}\big[R + 4(\nabla \Phi)^2\big] - \frac1{12} F_{mnp}F^{mnp}\Big]\ .
\end{equation}
The corresponding equations of motion and Bianchi identities are ($m,n, \ldots = 0,1,2,3,4,5$)
\begin{align}\nonumber
& R + 4 \nabla^2 \Phi - 4 (\nabla\Phi)^2 = 0\ ,
&& R_{mn} + 2 \nabla_m\nabla_n \Phi = \frac{e^{2\Phi}}4 (F_{mpq}F_n{}^{pq} - \frac16 g_{mn}F^2) \ ,
\\\label{seom3}
& \partial_p (\sqrt{-g}F^{mnp}) = 0 \ ,
&& \partial_{[q} F_{mnp]} = 0 \ .
\end{align}
The first two equations imply again that the dilaton should satisfy $\nabla^2 e^{-2 \Phi} =0$.

The metric of the \emo in this case is given by \ci{abf,hrt}
\begin{equation}\label{met3}
d\hat{s}_3^2 = -\frac{1+\rho^2}{1-\varkappa^2\rho^2} d\ot^2 + \frac{d\rho^2}{(1-\varkappa^2\rho^2)(1+\rho^2)} + \rho^2d\opsi_1^2
+\frac{1-r^2}{1+\varkappa^2r^2}d\ovarphi^2 + \frac{dr^2}{(1+\varkappa^2r^2)(1-r^2)} + r^2 d\ophi_1^2\ .
\end{equation}
It has a $U(1)^4$ isometry
represented by shifts in $\ot$, $\opsi_1$, $\ovarphi$ and $\ophi_1$
(we again use hats to denote isometric directions that we will T-dualize).

As in the $AdS_2 \times S^2$ case, the solution of \eqref{seom3} we find is for the metric \rf{met3}
T-dualized in all four isometric directions. The resulting background is
(the target-space indices
$m,n,\ldots = 0,1,2,3,4,5$ correspond to $ t,\psi_1,\r,\vp,\phi_1,r$)
\begin{align}\nonumber
&ds_3^2 = \eta_{ab}e^a e^b \ , \qquad
e^0 = \frac{\sqrt{1-\vk^2\rho^2}}{\sqrt{1+\rho^2}} dt \ , \qquad
e^1 = \frac{d\psi_1}{\rho} \ , \qquad
e^2 = \frac{d\rho}{\sqrt{1-\varkappa^2\rho^2}\sqrt{1+\rho^2}} \ , \qquad
\\ \nonumber
& \hspace{87pt}
e^3 = \frac{\sqrt{1+\vk^2r^2}}{\sqrt{1- r^2}} d\vp \ , \qquad
e^4 = \frac{d\phi_1}{r} \ , \qquad
e^5 = \frac{dr}{\sqrt{1+\varkappa^2r^2}\sqrt{1-r^2}} \ ,
\\ & e^{\Phi_3} F_3 = \frac{2i\sqrt{1+\vk^2}}{\sqrt{1-\vk^2\r^2}\sqrt{1+\vk^2r^2}}\nonumber
\big[(e^0 \W e^1 \W e^5 + e^2 \W e^3 \W e^4)
+
\varkappa^2 \rho r ( e^0 \W e^4 \W e^5 - e^1 \W e^2 \W e^3)
\\\nonumber & \hspace{140pt} + \varkappa \rho (e^0 \W e^3 \W e^4 - e^1 \W e^2 \W e^5) + \varkappa r (e^0 \W e^1 \W e^3 + e^2 \W e^4 \W e^5) \big]\ ,
\\\label{sol3}
& e^{\Phi_3} =
e^{\Phi_0 - 2\varkappa (t+\varphi )} \frac{(1-\varkappa^2\rho ^2) (1+\varkappa^2 r^2) }{\rho r\sqrt{1+ \rho ^2} \sqrt{1- r^2}}\ .
\end{align}
As in the \adt case in \rf{sol2}, \rf{n2} the RR flux takes very simple form when written in terms of the ``deformed'' basis introduced in \rf{n1}
\begin{align}\no
ds^2_3 & = \eta_{ab} e'^a e'^b \ , \qquad \qquad
e^{\Phi_3} F_3 = {2i\sqrt{1+\vk^2}}\,
\big( e'^0 \W e'^1 \W e'^5 + e'^2 \W e'^3 \W e'^4\big)\ ,
\\ e'^0 & = \eE^t \ , \qquad e'^1 =\eE^{\psi_1}\equiv e^1 \ , \qquad e'^2 = \eE^\r \ , \qquad e'^3= \eE^\vp \ , \qquad e'^4= \eE^{\phi_1} \equiv e^4 \ , \qquad e'^5 = \eE^r \ . \la{n3}
\end{align}
In this basis we again see that the $\vk$-deformation preserves the structure
of the undeformed
background; $e^{\P_3}F_3$ remains constant and is just rescaled by a factor
of $\sqrt{1+\vk^2}$.\foot{As written, the 3-form flux satisfies the 
self-duality equation
$F_3{}_{mnp} = \tfrac{1}{3!} \sqrt{-g}\, \e_{mnpqrs}F_3^{qrs}$,~$\e_{012345} = -1$.
As in the \adt case, there is actually a one-parameter family of fluxes that
solve the supergravity equations \eqref{seom3} given by
\begin{equation*}
e^{\Phi_3} F_3 = {2i\sqrt{1+\vk^2}}\,
\big(\text{c}_1\, e'^0 \W e'^1 \W e'^5 + \text{c}_2\, e'^2 \W e'^3 \W e'^4\big)\ ,
\qquad \text{c}_1^2 + \text{c}_2^2 = 2 \ .
\end{equation*}
Therefore, with an appropriate choice of $\text{c}_{1,2}$, the sign of the
self-duality equation can be reversed, for example, taking $\text{c}_1 = -1$,
$\text{c}_2 = 1$ ($\text{c}_1 = 1$, $\text{c}_2 = -1$). 
This can also be understood as reversing the sign of the isometric coordinate
$\psi_1$ ($\phi_1$). Indeed, reversing the sign of an odd number of the isometries 
of the metric will reverse the sign of the self-duality equation. However,
reversing the sign of $t$ or $\varphi$ will modify the dilaton as well as the
3-form flux.} 
Furthermore, the RR potential  $C_2$  for  $F_3 = d C_2$ again takes
a simple form in this basis (cf. \rf{28}) 
\begin{equation}\la{288}
e^{\Phi_3} C_2 = i\sqrt{1+\vk^{-2}}\Big(
\frac{e'^1 \wedge e'^5}{\sqrt{1+\r^2}} -
\frac{e'^2 \wedge e'^4}{\sqrt{1-r^2}}
\Big) \ .
\end{equation}

Some other important features of this solution are:
\begin{itemize}
\item The background fields entering the classical Green-Schwarz action (the metric and $e^\Phi F$) are invariant
under the $U(1)^4$ isometry given by shifts in $t$, $\psi_1$, $\varphi$ and $\phi_1$. This isometry is
broken to $U(1)^3$ in the dilaton by the linear $t+ \vp$ term.
\item For $\varkappa \in \mathbb{R}$ the metric and dilaton are real, while the RR flux is imaginary.
\item For $\varkappa \to 0$ the $U(1)^4$ isometry is restored in the full background, and
T-dualizing in all four isometries we recover
the standard solution for undeformed $AdS_3 \times S^3$ with constant dilaton and real homogeneous RR flux.
\item For $\varkappa \to \infty$, rescaling the fields and the string tension, we find a non-standard background,
i.e. the dilaton still has a linear dependence on the isometric coordinates.
\item For $\varkappa \to i$ the 3-form flux vanishes, while the $t$ and $\varphi$ directions become free
in the metric. We will discuss this limit in more detail in section \ref{kai}.
\end{itemize}

As in the $AdS_2 \times S^2$ case we may formally T-dualize 
the background \eqref{sol3} in all four isometries of the metric
to recover the metric in \eqref{met3}. The resulting dilaton will then
be (cf. \rf{di2})
\begin{equation}
e^{\hat \Phi_3} = e^{\Phi_0 - 2\varkappa(t+\varphi)}\sqrt{1-\vk^2\r^2}\sqrt{1+\vk^2 r^2}\ ,
\end{equation}
i.e. linear in the dual coordinates and constant in the $\vk \to 0$ limit.

Again, there is a second solution we can find -- directly for the metric \eqref{met3}
with no T-dualities applied. This solution is related to \eqref{sol3} by the
formal map\foot{If we also analytically continue $\vk$ to the region
$\varkappa^2 \in (0,-1]$ including the point $\varkappa = i$ then
the map \eqref{map3} is a real diffeomorphism.}
\begin{equation}\label{map3}
t \to \frac{i \ot}{\varkappa} \ , \qquad \rho \to \frac{i}{\varkappa\rho} \ , \qquad \psi_1 \to \frac{i \opsi_1}{\varkappa} \ , \qquad
\varphi \to \frac{i \ovarphi}{\varkappa} \ , \qquad r \to \frac{i}{\varkappa r} \ , \qquad \phi_1 \to \frac{i\ophi_1}{\varkappa} \ .
\end{equation}
However, the $\varkappa \to 0$ limit of the resulting background gives a
non-standard solution -- the undeformed $AdS_3 \times S^3$ metric supported by a complex
dilaton (with a linear dependence on the isometric directions) and complex RR
flux. But it does have a natural $\varkappa \to \infty$ limit if we first
use the rescaling
\begin{equation}\begin{split}\label{kainflim3}
& \ot \to \varkappa^{-1} \ot \ , \qquad
\rho \to \varkappa^{-1} \rho \ , \qquad
\ovarphi \to \varkappa^{-1} \ovarphi \ , \qquad
r \to \varkappa^{-1} r \ ,
\\
& ds^2 \to \vk^2 ds^2 \ , \qquad e^{\Phi_3} F_3 \to \vk^2 e^{\Phi_3} F_3 \ , \qquad e^{\Phi_3} \to \vk^{-4} e^{\Phi_3} \ ,
\end{split}\end{equation}
along with a shift of the constant
part of the dilaton $\Phi_0$ by $\frac{i\pi}2$. The resulting real background is the ``mirror''
model one \cite{mirror0} and is related to a $dS_3 \times H^3$
background by T-dualities in $\ot$ and $\ovarphi$,
with a constant dilaton and imaginary homogeneous RR flux (as could be expected).
In this sense expanding around $\varkappa = \infty$ corresponds
to considering the \emo for $dS_3 \times H^3$ with deformation parameter
$\widetilde \vk = \vk^{-1}$.

\subsection{\texpdf{$AdS_5 \times S^5$}{AdS5 x S5}}\label{secsugra5}

Let us now turn to the case of the $\eta$-deformed $AdS_5 \times S^5$.
The background that we find can be interpreted as a deformation of the T-dual of \adss
and
is a solution of 10d type IIB supergravity with just the metric, dilaton
and RR 5-form flux switched on.\foot{Applying T-duality in all 6 isometric
directions of undeformed \adss metric one gets a formal
type IIB solution with T-dual metric supported by
non-constant dilaton and (imaginary) 5-form flux.
The reason for finding only non-zero RR 5-form flux
can be understood heuristically by observing that T-duality is applied to the time and 2
longitudinal directions (angles of $S^3$) of the D3-branes as well as 3 transverse directions (3 angles of the transverse $S^5$) with the longitudinal and transverse directions interchanging their roles. }

The relevant part of the type II 10d supergravity action is then given by
\begin{equation}\label{sact5}
\mathcal{S}_{10} = \int d^{10}x \; \sqrt{-g}\Big[e^{-2\Phi}\big[R + 4(\nabla \Phi)^2\big] - \frac1{4\cdot 5!} F_{mnpqr}F^{mnpqr}\Big]\ .
\end{equation}
The corresponding equations of motion and Bianchi identities are ($m,n, \ldots = 0,1,2,3,4,5,6,7,8,9$)
\begin{align}\nonumber
& R + 4 \nabla^2 \Phi - 4 (\nabla\Phi)^2 = 0\ ,
&& R_{mn} + 2 \nabla_m\nabla_n \Phi = \frac{e^{2\Phi}}{4\cdot 4!} F_{mpqrs}F_n{}^{pqrs} \ ,
\\\label{seom5}
& \partial_r (\sqrt{-g}F^{mnpqr}) = 0 \ ,
&&
\!\!\!\!\!\!\!\!\!\!\!\!\!
\partial_{[s} F_{mnpqr]} = 0 \ ,
\qquad F_{mnpqr} = -\frac{1}{5!}\sqrt{-g}\, \epsilon_{mnpqrstuvw}F^{stuvw} \ ,
\end{align}
where we have also included the self-duality equation ($\epsilon_{0123456789} = -1$)
for the RR 5-form flux, which needs to be imposed separately.
Again, the first two equations imply that the dilaton should satisfy $\nabla^2 e^{-2 \Phi} =0$.

The metric and $B$-field corresponding to the
\emo in this case are \ci{abf}
(as before, we use hats to denote isometries that we will T-dualize)
\begin{align}\nonumber
d\hat{s}_5^2 & = -\frac{1+\rho^2}{1-\varkappa^2\rho^2} d\ot^2 + \frac{d\rho^2}{(1-\varkappa^2\rho^2)(1+\rho^2)} + \frac{\rho^2\cos^2\zeta}{1+\varkappa^2 \rho^4\sin^2\zeta}d\opsi_1^2 + \frac{d\zeta^2}{1+\varkappa^2 \rho^4 \sin^2 \zeta} +\rho^2 \sin^2\zeta \, d\opsi_2^2
\\\nonumber & \hspace{11pt}+\frac{1-r^2}{1+\varkappa^2r^2}d\ovarphi^2 + \frac{dr^2}{(1+\varkappa^2r^2)(1-r^2)} + \frac{r^2 \cos^2\xi}{1+\varkappa^2 r^4 \sin^2 \xi} d\ophi_1^2 + \frac{d\xi^2}{1+\varkappa^2 r^4 \sin^2 \xi} + r^2 \sin^2\xi\, d\ophi_2^2\ ,
\\\label{met5}
\hat B & = \frac{\varkappa \rho^4 \sin \zeta\cos\zeta}{1+\varkappa^2 \rho^4 \sin^2\zeta} d\opsi_1 \W d\zeta - \frac{\varkappa r^4 \sin \xi \cos \xi}{1+\varkappa^2 r^4 \sin^2\xi} d\ophi_1 \W d \xi \ . 
\end{align}
Both have a $U(1)^6$ isometry
(corresponding to the Cartan directions
of the bosonic subgroup of the undeformed $PSU(2,2|4)$ symmetry)
represented by
shifts in $\ot$, $\opsi_1$, $\opsi_2$, $\ovarphi$, $\ophi_1$ and $\ophi_2$.

The type IIB solution supporting {\it that} metric is not known (cf. \ci{abf2})
but as in the lower-dimensional examples above we will find a consistent solution that supports the fully
T-dual metric and $B$-field, i.e. the background
\rf{met5}
T-dualized in all six isometries
\begin{align}
ds^2_5=\, &- { 1-\varkappa^2\rho^2 \ov 1+\rho^2}dt^2
+\frac{d\rho^2}{(1+\rho^2) (1-\varkappa^2\rho^2) }
\no \\ & \qquad
+ \frac{ 1+\varkappa^2\rho^4\sin^2\zeta }{\rho^2\cos^2\zeta} \Big(d\psi_1 {\,\reo +\,}
{\varkappa \rho^4 \sin \zeta\, \cos \zeta \ov 1+\varkappa^2\rho^4\sin^2\zeta } d\zeta\Big)^2
+{ \rho^2 d\zeta^2 \ov 1+\varkappa^2\rho^4\sin^2\zeta}
+{d\psi_2^2 \ov \rho^2\sin^2\zeta } \no \\
&+ { 1+\varkappa^2 r^2 \ov 1- r^2}d\vp ^2
+\frac{d r^2}{(1-r^2) (1+\varkappa^2 r^2) }
\no \\ & \qquad \nonumber
+ \frac{ 1+\varkappa^2r^4\sin^2\xi }{r^2\cos^2\xi} \Big(d\phi_1 {\,\reo -\,}
{\varkappa r^4 \sin \xi\, \cos \xi \ov 1+\varkappa^2r^4\sin^2\xi } d\xi\Big)^2
+{ r^2 d\xi^2 \ov 1+\varkappa^2 r^4\sin^2\xi}
+{d\phi_2^2 \ov r^2\sin^2\xi }
\\ =\, & - { 1-\varkappa^2\rho^2 \ov 1+\rho^2}dt^2
+\frac{d\rho^2}{(1+\rho^2) (1-\varkappa^2\rho^2) }
+{d\psi_1^2 \ov \rho^2\cos^2\zeta } \no
+ (\r\, d\z + \vk \r \tan\z\, d\psi_1)^2
+{d\psi_2^2 \ov \rho^2\sin^2\zeta } \\
&+ { 1+\varkappa^2 r^2 \ov 1- r^2}d\vp ^2
+\frac{d r^2}{(1-r^2) (1+\varkappa^2 r^2) }
+{d\phi_1^2 \ov r^2\cos^2\xi }
+ (r\, d\xi - \vk r \tan\xi\, d\phi_1)^2
+{d\phi_2^2 \ov r^2\sin^2\xi } \ , \la{m5}
\\ B =\, & 0 \ . \la{b5}
\end{align}
Here the T-dualities in $\opsi_1$ and $\ophi_1$
removed the $B$-field at the expense of introducing off-diagonal terms in
the metric as in \ci{lrt}.
The T-duality conventions we use are 
given in Appendix \ref{appa}.

In \rf{m5} we have presented two forms of the dualized metric;
the first is the one that arises naturally from the T-duality procedure
\cite{lrt}, while the second has a particularly simple form, which will
be useful later.

Introducing  the shorthand  notation 
$E^{abcde} \equiv  e^a \W e^b \W e^c \W e^d \W e^e$, 
and  
taking the target-space indices $m,n,\ldots = 0,1,2,3,4,5,6,7,8,9$
to correspond to $t,\psi_2,\psi_1,\z,\r,\vp,\phi_2,\phi_1,\xi,r$, the
solution of \eqref{seom5} we found is
\small
{\allowdisplaybreaks
\begin{align}\nonumber
&ds_5^2 = \eta_{ab}e^a e^b \ , \qquad
e^0 = \frac{\sqrt{1-\vk^2\rho^2}}{\sqrt{1+\rho^2}} dt \ , \qquad
e^1 = \frac{d\psi_2}{\rho\sin\zeta}\ ,\qquad
e^2 = \frac{\sqrt{1+\varkappa^2\rho^4\sin^2\zeta}}{\rho \cos\zeta}d\psi_1 {\,\reo +\,} \frac{\varkappa\rho^3\sin\zeta}{\sqrt{1+\varkappa^2\rho^4\sin^2\zeta}}d\zeta \ ,
\\ \nonumber & \hspace{81pt}
e^3 = \frac{\rho\,d\zeta}{\sqrt{1+\varkappa^2\rho^4\sin^2\zeta}} \ ,\qquad
e^4 = \frac{d\rho}{\sqrt{1-\varkappa^2\rho^2}\sqrt{1+\rho^2}} \ ,
\\ \nonumber
& \hspace{81pt}
e^5 = \frac{\sqrt{1+\vk^2r^2}}{\sqrt{1- r^2}} d\vp \ , \qquad
e^6 = \frac{d\phi_2}{r\sin \xi} \ , \qquad
e^7 = \frac{\sqrt{1+\varkappa^2r^4\sin^2\xi}}{r\cos\xi}d\phi_1 {\,\reo -\,} \frac{\varkappa r^3 \sin\xi}{\sqrt{1+\varkappa^2r^4 \sin^2\xi}}d\xi \ ,
\\ \nonumber & \hspace{81pt}
e^8 = \frac{r\,d\xi}{\sqrt{1+\varkappa^2r^4 \sin^2\xi}}\ , \qquad
e^9 = \frac{dr}{\sqrt{1+\varkappa^2r^2}\sqrt{1-r^2}} \ ,
\\ \nonumber &
\\ \nonumber & e^{\Phi_5} F_5 = \frac{4i\sqrt{1+\vk^2}}{\sqrt{1-\vk^2\r^2}\sqrt{1+\vk^2\r^4\sin^2\zeta}\sqrt{1+\vk^2r^2}\sqrt{1+\vk^2\rho^4\sin^2\xi}}\nonumber
\\\nonumber & \hspace{74pt} \Big[\big[
E^{01289} + E^{34567}
- \vk \r (E^{03567} - E^{12489})
- \vk r (E^{01258} + E^{34679})
+ \vk^2 \r r (E^{03679} - E^{12458})\big]
\\\nonumber & \hspace{35pt}
{\reo -\,} \vk \r^2 \sin \z \big[ E^{01389} - E^{24567}
+ \vk \r (E^{02567} + E^{13489})
- \vk r (E^{01358} - E^{24679})
- \vk^2 \r r (E^{02679} + E^{13458}) \big]
\\\nonumber & \hspace{35pt}
{\reo -\,} \vk r^2 \sin \xi \big[ E^{01279} - E^{34568}
+ \vk \r (E^{03568} + E^{12479})
- \vk r (E^{01257} - E^{34689})
- \vk^2 \r r (E^{03689} + E^{12457}) \big]
\\\nonumber &
+ \vk^2 \r^2 r^2\sin \z\sin \xi \big[E^{01379} + E^{24568}
- \vk \r (E^{02568} - E^{13479})
- \vk r (E^{01357} + E^{24689})
+ \vk^2 \r r (E^{02689} - E^{13457}) \big] \Big]
\\\label{sol5}
& e^{\Phi_5} =
e^{\Phi_0 - 4\varkappa (t+\varphi ) {\reo -} 2 \vk (\psi_1 - \phi_1)} \frac{(1-\varkappa^2\rho ^2)^2 (1+\varkappa^2 r^2)^2 }{\rho^2 r^2\sqrt{1+ \rho ^2} \sqrt{1- r^2}\sin2\z \sin 2\xi}\ .
\end{align}}\normalsize
Remarkably, as in the lower-dimensional cases in \rf{n2}, \rf{n3},
this complicated-looking RR flux once again
takes a very simple form when written in terms of the ``deformed'' basis in \rf{n1}
and the angular part of the vielbein associated with the second form of the metric
in \rf{m5}
\begin{align}
ds_5^2 & = \eta_{ab}e'^a e'^b \ , \qquad 
e^{\Phi_5} F_5& = {4i\sqrt{1+\vk^2}}
\big(
e'^0 \W e'^1 \W e'^2 \W e'^8 \W e'^9 + e'^3 \W e'^4 \W e'^5 \W e'^6 \W e'^7 \big) \ , \la{n5} \end{align}
where
\be
&& \no
e'^0= \eE^t \ , \qquad\quad \hspace{3pt} e'^1 = \eE^{\psi_2} \equiv e^1 \ , \qquad\hspace{9pt}
e'^2= \eE^{\psi_1} \ , \qquad\quad
e'^3 = \eE^\zeta \ , \qquad\hspace{9pt} e'^4 = \eE^\r \ ,
\\\no
&&
e'^5 = \eE^\vp \ , \qquad\quad
e'^6 = \eE^{\phi_2} \equiv e^6 \ , \qquad\quad
e'^7= \eE^{\phi_1}\ , \qquad\quad e'^8= \eE^{\xi}\ ,
\qquad\quad e'^9= \eE^r \ , \\
&&\no
\eE^{\psi_1} \equiv { e^2 -\vk \r^2 \sin \z\, e^3 \ov \sqrt{1+\vk^2\r^4\sin^2\zeta} }
={d \psi_1 \ov \r \cos \z} \ , \ \ \ \ \
\eE^\zeta \equiv { e^3 + \vk \r^2 \sin \z\, e^2 \ov \sqrt{1+\vk^2\r^4\sin^2\zeta} }
= \r d\z + \vk \r \tan \z \, d\psi_1
\ , \\
&&\la{pp5}
\eE^{\phi_1} \equiv { e^7 + \vk r^2 \sin \xi\, e^8 \ov \sqrt{1+\vk^2r^4\sin^2\xi} }
={d \phi_1 \ov r \cos \xi} \ , \ \ \ \ \ 
\eE^\xi \equiv { e^8 - \vk r^2 \sin \xi\, e^7 \ov \sqrt{1+\vk^2r^4\sin^2\xi} }
= r d\xi - \vk r \tan \xi \, d\phi_1
\ .
\ee
In this basis we again see that the $\vk$-deformation preserves the structure
of the undeformed background; $e^{\P_5}F_5$ remains constant and is just
rescaled by a factor of $\sqrt{1+\vk^2}$.\foot{By construction, the 5-form flux
satisfies the 
self-duality equation given in \eqref{seom5}.
If we drop the requirement of self-duality,
there are a discrete set of four fluxes that
solve the supergravity equations \eqref{seom5} given by
\begin{equation*}
e^{\Phi_5} F_5 = {4i\sqrt{1+\vk^2}}
\big(
\text{c}_1\,e'^0 \W e'^1 \W e'^2 \W e'^8 \W e'^9 +
\text{c}_2\, e'^3 \W e'^4 \W e'^5 \W e'^6 \W e'^7 \big) \ ,
\qquad \text{c}_1^2 = \text{c}_2^2=1 \ .
\end{equation*}
This is in contrast to the lower-dimensional cases for which there was a
one-parameter family of fluxes. However, as in those cases, with an
appropriate choice of $\text{c}_{1,2}$ the sign of the self-duality equation
can still be reversed, for example, taking $\text{c}_1 = -1$, $\text{c}_2 =
1$ ($\text{c}_1 = 1$, $\text{c}_2 = -1$). This can also be understood as
reversing the sign of the isometric coordinate $\psi_2$ ($\phi_2$). Indeed,
reversing the sign of an odd number of the isometries of the metric will
reverse the sign of the self-duality equation. However,
reversing the sign of $t$, $\psi_1$, $\varphi$ or $\phi_1$ will modify the metric and dilaton
as well as the 5-form flux.}
Furthermore, as in the lower-dimensional cases \rf{28} and \rf{288}, 
the RR potential $C_4$ for $F_5 = d C_4$ takes a simple form in this basis
\begin{equation}
e^{\Phi_5} C_4 =  i\sqrt{1+\vk^{-2}}\Big(
\frac{e'^1 \wedge e'^2 \wedge e'^8 \wedge e'^9}{\sqrt{1+\r^2}} +
\frac{e'^3 \wedge e'^4 \wedge e'^6 \wedge e'^7}{\sqrt{1-r^2}}
\Big) \ .
\end{equation}
The singular part of $C_4$ in the $\vk\to 0$ limit is again pure gauge. 

Some important features of this solution are:
\begin{itemize}
\item The background fields entering the classical Green-Schwarz action (the metric and $e^\Phi F_5$) are invariant
under the $U(1)^6$ isometry given by shifts in $t$, $\psi_1$, $\psi_2$, $\varphi$, $\phi_1$ and $\phi_2$. This isometry is
broken to $U(1)^5$
by the presence of one linear combination of
four of these directions in the dilaton $\Phi_5$, and hence the isometry is also broken in $F_5$.

\item For $\varkappa \in \mathbb{R}$ the metric and dilaton are real, while the RR flux is imaginary
(which may be attributed to T-duality in time direction being secretly involved).
\item For $\varkappa \to 0$ the $U(1)^6$ isometry is restored in the full background, and we
can T-dualize back to the frame of \rf{met5}.
T-dualizing in all six isometries
we recover
the standard solution for $AdS_5 \times S^5$ (i.e. the $\varkappa\to 0$ limit of \rf{met5}) with constant dilaton and real
homogeneous RR flux.
\item For $\varkappa \to \infty$, rescaling the fields and the string tension, we find a non-standard background, i.e. the
dilaton still has a linear dependence on the isometric coordinates.
\item For $\varkappa \to i$ the RR flux vanishes, while the $t$ and $\varphi$ directions become free.
We will discuss this limit in more detail in section \ref{kai}.
\end{itemize}

If we were allowed to T-dualize the background \rf{sol5} in all six isometries of the metric
using the standard rules and ignoring the linear terms in the dilaton,
we would end up with the metric and $B$-field in \rf{met5} and a RR background $e^{\hat \Phi_5} \hat F_n$,
{which is precisely the one extracted from the quadratic fermionic term
of the deformed supercoset action in \ci{abf2} (see Appendix \ref{appa}).}
The dilaton formally
corrected by the standard factor originating from T-duality will then be (cf. \rf{di2})
\be
\la{di5}
e^{\hat \Phi_5} =
e^{\Phi_0 - 4\vk (t+\varphi ) {\reo -} 2 \vk (\psi_1 - \phi_1)}
\Big[\frac{(1-\varkappa^2\rho ^2)^3(1+\varkappa^2 r^2)^3}
{ ({1+ \varkappa^2 \rho ^4\sin^2 \zeta}) ( {1 + \varkappa^2 r ^4\sin^2 \xi} )}\Big]^{1/2} \ , 
\ee
i.e. will be constant in the $\vk\to 0$ limit.
Once again, the resulting background
will not have an interpretation as a standard type IIB solution as
the dilaton (and thus the RR flux) will depend on the original isometric coordinates while the metric will
describe
their dual counterparts.

As in the lower-dimensional cases there is formally a second solution
we can find, which is
for the metric and $B$-field in \eqref{met5} T-dualized {\it only} in $\opsi_1$ and $\ophi_1$
(i.e. obtained by doing only the T-dualities that remove the $B$-field at the expense of introducing
off-diagonal terms in the metric).
This T-duality turns out to be necessary --
in contrast to the lower-dimensional \adt and \adst cases, here we still are unable to find
(even a formal, complex)
type IIB solution
supporting directly the original metric and $B$-field in \rf{met5}.
This solution is related to \eqref{sol5} by the map
\begin{equation}\begin{split}\label{map5}
& t \to \frac{i \ot}{\varkappa} \ , \qquad \ \, \rho \to \frac{i}{\varkappa\rho} \ , \qquad \psi_1 \to \psi_1 {\,\reo +\,} \frac{1}{\varkappa} \log \sin \zeta \ , \qquad
\z \to i \log \tan \frac{\zeta}2 + \frac{\pi}2 \ , \qquad \psi_2 \to \frac{i\opsi_2}{\varkappa} \ ,
\\ & \varphi \to \frac{i \ovarphi}{\varkappa} \ , \qquad r \to \frac{i}{\varkappa r} \ , \qquad \phi_1 \to \phi_1 {\,\reo -\,} \frac{1}{\varkappa}\log \sin\xi \ , \qquad
\, \xi \to i \log\tan \frac{\xi}2 + \frac{\pi}2 \ , \qquad \phi_2 \to \frac{i\ophi_2}{\varkappa} \ .
\end{split}\end{equation}
The $\varkappa \to 0$ limit of the resulting background is an unfamiliar solution representing
the undeformed $AdS_5 \times S^5$ metric supported by a complex
dilaton with a linear dependence on the isometric directions and complex RR
flux. It does, however, have a natural $\varkappa \to \infty$ limit if we first
do the rescaling
\begin{equation}\begin{split}\label{kainflim5}
& \ot \to \varkappa^{-1} \ot \ , \qquad \ \,
\rho \to \varkappa^{-1} \rho \ , \qquad
\psi_1 \to \varkappa^{-2} \psi_1 \ , \qquad
\\
& \ovarphi \to \varkappa^{-1} \ovarphi \ , \qquad
r \to \varkappa^{-1} r \ , \qquad
\phi_1 \to \varkappa^{-2} \phi_1 \ , \qquad
\\
& ds^2 \to \vk^2 ds^2 \ , \qquad e^{\Phi_5} F_5 \to \vk^4 e^{\Phi_5} F_5 \ , \qquad e^{\Phi_5} \to \vk^{-8} e^{\Phi_5} \ ,
\end{split}\end{equation}
along with a shift of the constant part of the dilaton $\Phi_0$ by $\frac{i\pi}2$.
T-dualizing the resulting real background in $\psi_1$ and $\phi_1$ we find the ``mirror''
model solution constructed in \cite{mirror0}, which is furthermore related to a $dS_5 \times H^5$
background
(by T-dualities in $\ot$ and $\ovarphi$)
with constant dilaton and imaginary homogeneous RR 5-form flux.
Thus considering $\vk$ in the region around $\varkappa = \infty$ corresponds
to the \emo for $dS_5 \times H^5$ with the deformation parameter
$\widetilde \vk = \vk^{-1}$.

\section{\texpdf{$\vk \to i$}{kappa -> i} limit}\label{kai}

In this section we will consider the $\eta \to i$, or, equivalently,
$\varkappa \to i$ (cf. \rf{01}) limit of the above backgrounds.\foot{This limit will be considered formally, i.e.
the resulting background may be complex and one may need further analytic continuation to get a real solution.}

This limit can be taken in two different ways.
Setting $\vk=i$ directly (with all coordinates fixed) one finds that the RR flux vanishes, the two
``longitudinal'' coordinates $t$ and $\vp$ decouple, and the
resulting ``transverse'' metric-dilaton background
can be interpreted as corresponding to a special limit of a gauged WZW model.

Alternatively, the limit can be taken by
combining $\vk \to i$ with a special rescaling of the coordinates $x^\pm = t \mp \vp$ which leads
to a pp-wave type background \ci{hrt}.
As was found in \ci{hrt}, for the $AdS_n \times S^n$ cases with $n=2,3$,
fixing the light-cone gauge in the string action for the
pp-wave background one arrives at the Pohlmeyer-reduced (PR) model \cite{gt} for the corresponding
undeformed $AdS_n \times S^n$ model.\foot{In \ci{hrt} the corresponding pp-wave RR background for the cases $n=2,3$
was reconstructed directly in the pp-wave limit, while here we will derive it from the full deformed solution.}
In the $n=5$ case we shall show that
a similar procedure leads to a special limit \ci{ht} of the PR model for the $AdS_5 \times S^5$
superstring.

\subsection{Direct \texpdf{$\vk\to i$}{kappa -> i} limit and connection with the gauged WZW model}\label{3.1}

Setting $\vk=i$
in the backgrounds \eqref{sol2}, \eqref{sol3} and
\eqref{sol5} we find the following common structure of the corresponding
metric, dilaton and RR fluxes ($n=2,3,5$)
\be \la{1}
ds_n^2 = -dt^2 + d\vp^2 + ds_{n\perp}^2 \ , \qquad
\Phi_n = \Phi_{n \parallel} + \Phi_{n \perp} \ , \qquad \Phi_{n \parallel} = - i(n-1) (t + \vp) \ , \qquad
F_n =0 \ .
\ee
First, as the RR fluxes in \eqref{sol2}, \eqref{sol3}, \eqref{sol5} all have an overall factor of $\sqrt{1+\vk^2}$ they
vanish for $\vk = i$.
As a result, we get a purely NS-NS metric-dilaton background that must be a solution
of the supergravity equations.
Second, the ``longitudinal'' $t$ and $\vp$ directions effectively decouple from the remaining
transverse directions (they form an
$\mathbb{R}^{1,1}$ subspace in the metric). The ``null'' linear dilaton term $\Phi_{n \parallel} $ then
does not affect the value of the
central charge, i.e. one also gets a solution when taking this term
with an arbitrary coefficient.\foot{This dilaton can be made real by an analytic continuation
interchanging the roles of $t$ and $\vp$, i.e. $t= i \vp', \ \ \vp = i t'$. }

Thus the transverse metric and dilaton should represent a conformal sigma model on their own.
Since there is no RR flux and the metric and dilaton have a direct-product $M^{n-1}_A \times M^{n-1}_S$ structure,
we should end up
with the direct sum of the two transverse conformal models corresponding to the $AdS_n$ and $S^n$ parts of the
deformed background, i.e. having
$(n-1)$-dimensional target spaces ($n=2,3,5$).\foot{One can also formally consider the
3d model corresponding to the $n=4$ case (i.e. $AdS_4 \times S^4$), which
can be viewed as a truncation of the $n=5$ case.}

Explicitly, setting $\vk =i$ in \eqref{sol2} and \eqref{sol3} and redefining the ``radial'' coordinates as
\begin{equation}
\rho \equiv \tan \alpha \ , \qquad \qquad r \equiv \tanh \beta \ ,\la{rrr}
\end{equation}
we find that for $n=2 $ and $n=3$
\begin{align}
& ds_{2\perp}^2 = d\a^2 + d\b^2 \ , &&
{\Phi_{2\perp}} = {\Phi_0} \ , \la{d2}
\\
& ds_{3\perp}^2 = d\a^2 + { \cot^2 \a\, d\psi_1^2} + d\b^2 + { \coth^2 \b\, d\phi_1^2 } \ ,
&& e^{\Phi_{3\perp} } = { e^{\Phi_0} \ov \sin \alpha \ \sinh \beta} \ .\la{d3}
\end{align}
Thus in the $AdS_2 \times S^2$ case we find a free transverse theory, while in the $AdS_3 \times S^3$ case
it is represented by direct sum of the (vector) gauged WZW models
for $G/H= SO(3)/SO(2)$ and $SO(1,2)/SO(2)$ \ci{rabi}.
In the $n=5$ case we get from \rf{sol5}
\begin{align}
& ds_{5\perp}^2 = ds_{5A\perp}^2 + ds_{ 5S\perp}^2 \ , \ \ \ \ \ \ \ \ \ \ \
\Phi_{5\perp} =\Phi_0 + \Phi_{5A\perp} + \Phi_{5S\perp} \ , \la{34} \\
&ds_{5A\perp}^2= d\a^2 + \cot^2 \a \, \Big({d\psi_1^2 \ov \cos^2\z} + {d\psi_2^2 \ov \sin^2\z}\Big)
+ \tan^2 \a \, (d\z {\,\reo +\,} i \tan \z\, d\psi_1)^2 \ , \no \\
&e^{\Phi_{5A\perp}} =
{e^{ {\reo -} 2 i \psi_1 } \ov \cos \a\, \sin^2 \a\, \sin2\z }\ , \la{d5}
\\ & ds_{5S\perp}^2= d\b^2 + \coth^2 \b \,\Big({d\phi_1^2 \ov \cos^2\xi} + {d\phi_2^2 \ov \sin^2\xi}\Big)
+ \tanh^2 \b \, (d\xi {\,\reo -\,} i \tan \xi\, d\phi_1)^2 \ , \no \\
&
e^{\Phi_{5S\perp}} =
{e^{2 i \phi_1} \ov \cosh \b\, \sinh^2 \b\, \sin 2\xi} \ . \la{d5s}
\end{align}
As for the $ [SO(3)/SO(2) ] \times [SO(1,2)/SO(2)]$ gauged WZW model in the $n=3$ case \rf{d3} above,
one can check directly that each of these two 4d metric-dilaton backgrounds satisfies
the corresponding equations in \rf{seom5} with the constant shifts of the central charge from the free-theory value, $4$,
cancelling between the two factors.
Note that the metrics and dilatons here can be made real by the analytic continuation
$
\psi_1 \to i \psi_1 , \ \phi_1 \to i \phi_1 $.

The gauged WZW interpretation of the two 4d backgrounds in \rf{d5}, \rf{d5s} may seem doubtful at first as here
each factor has 2 isometries, while the metrics corresponding to $G/H$ gauged WZW models with non-abelian $H$ should have no
isometries.
However, the isometries can be effectively generated by taking special singular
limits as pointed out in \ci{ht}.
Indeed, as we shall now explain, these two 4d backgrounds may be viewed as (an analytic continuation of) a singular limit
of those associated to the $SO(5)/SO(4)$ and $SO(1,4)/SO(4)$ gauged WZW models.

Let us start with the
metric-dilaton background corresponding to the $ SO(5)/SO(4)$ gauged WZW model as given in \ci{bs,rt}\foot{This solution can
be found by analytically continuing the result of \ci{bs} in the $\epsilon = \epsilon'=1$ coordinate patch, which is different to the
patch used to obtain the metric given in \cite{rt}.
The notation we use for the four angular coordinates is chosen for comparison with \rf{d5}.}
\begin{align}
& ds^2_{\rm gwzw} = d\a^2 + \cot^2 \a \, \big( {d\tpsi_1^2\ov \cos^2 \z }
+{ d \psi_2^2\ov \sin^2 \z }\big)
+ \tan^2 \a\ \Big( d \z - { \tan \z \, \sin 2\tpsi_1 \ d\tpsi_1 - \cot \z \, \sin 2 \psi_2 \, d \psi_2 \ov \cos 2\tpsi_1 + \cos 2 \psi_2 } \Big )^2 \ , \la{38} \\
&
e^{\Phi_{\rm gwzw}} = { 1 \ov \cos \a \, \sin ^ 2\a \, \sin 2 \z\, (\cos 2\tpsi_1 + \cos 2 \psi_2 ) } \ . \la{39}
\end{align}
We now consider the following singular limit (a special case of the limit
in \ci{ht}).
We first analytically continue $\tpsi_1\to i \vartheta $ and
shift $\vartheta$ by an infinite constant $L$,
i.e. set
\be \tpsi_1 = i(\vartheta + L)\ , \ \ \ \qquad L \to \infty \ . \ee
As a result, $\vartheta$ becomes an isometry of the metric \rf{38} while the dilaton (with $\P_0 = \P_0' + 2 L - \log 2$)
now has a linear dependence on $\vartheta$
\begin{align}
& ds'^2_{\rm gwzw} = d\a^2 + \cot^2 \a\, \big( - {d\vartheta^2\ov \cos^2 \zeta } +{ d \psi^2_2\ov \sin^2 \zeta }\big)
+ \tan^2 \a\ \big( d \z + { \tan \z\, d\vartheta } \big )^2 \ , \la{388} \\
&
e^{\Phi'_{\rm gwzw}} = { e^{ - 2 \vartheta } \ov \cos \a \, \sin ^ 2\a \, \sin 2 \z\ } \ . \la{399}
\end{align}
After taking the limit the coordinate $\psi_2$ also becomes an isometry of the full background
(this is a special feature of the limits discussed in \ci{ht}).
Comparing \rf{388}, \rf{399} to the metric and dilaton in \rf{d5} we see that analytically continuing back, $\vartheta = i\psi_1$, we recover
the $AdS_5$ part of the background \rf{34} found from the $\vk \to i$ limit of \eqref{sol5}.

Furthermore, we can start from the $SO(1,4)/SO(4)$ gauged WZW model, the metric and dilaton of which can be found by analytically continuing \eqref{38}, \eqref{39} as
\begin{equation}
\alpha \to i \beta \ , \qquad \tpsi_1 \to \tilde \phi_1 \ ,
\qquad
\psi_2 \to \phi_2 \ , \qquad \zeta \to \xi \ , \qquad ds^2 \to - ds^2 \ .
\end{equation}
Then taking a similar limit we recover the $S^5$ part of the background
\eqref{d5s}
found from the $\vk \to i$ limit of \eqref{sol5} (up to a constant shift of the dilaton).

Let us note that
these metrics
can be put into simple diagonal forms
where the shift isometry becomes a rescaling symmetry.
For example, introducing
$w=- \cos 2 \psi_2$, $v= \cos 2 \tpsi_1$, $u= \sin^2 \z \cos 2 \tpsi_1- \cos^2 \z \cos 2 \psi_2 $, one may write \rf{38} as \ci{bs}
\be
ds^2_{\rm gwzw} =
d\a^2 + \cot^2 \a\, \Big[ { (v-w) \, d w^2 \ov 4 ( u- w) (1-w^2) }
+ { (v-w)\, d v^2 \ov 4 (v - u) (1-v^2) } \Big] + \tan^2 \a \, { d u^2 \ov 4 ( u- w) (v - u) }
\ . \la{10}
\ee
The special limit described above then translates into an infinite rescaling
of the two coordinates
$v = e^{2L} v' , \ u =e^{2L} u'$.
This
produces the diagonal form of the background \rf{388}, \rf{399}
\ba
&&ds'^2_{\rm gwzw} =
d\a^2
+{ \cot^2 \a\ dx^2 - \tan^2 \a \ d y^2 \ov y^2-x^2 }+ \cot^2 \a\, { x^2 \, \ov y^2} \, d\psi_2^2
\ , \la{101}\\
&& e^{\Phi'_{\rm gwzw} }= { 1 \ov 2 \cos \a\, \sin^2\a \, y \sqrt{ x^2-y^2 }} \ , \qquad \qquad
x= \sqrt{ v'} = e^\vartheta \ , \ \ \ \ y= \sqrt{ u'} = \sin \z\, e^{\vartheta} \ . \la{111}
\eea
The two isometries of the metric \rf{101} are $(i)$ the simultaneous rescaling of $x$ and $y$ and $(ii)$ the
shift of $\psi_2$.\foot{Note that this metric is a direct generalization of the metric that appears upon taking
a scaling limit ($p\to e^L p, \ q\to e^L q , \ L\to \infty$) in the $SO(4)/SO(3)$ gauged WZW metric \ci{fl}
$ds^2 = d\a^2 +{ \cot^2 \a \ d p^2 + \tan^2 \a\ dq^2 \ov 1- p^2 - q^2 }$
with $p= x, \ q= i y$.}

We have thus demonstrated that the $\vk\to i$ limit of the $\vk$-deformed solution we have found
has an interpretation as a limiting background for the standard bosonic gauged WZW model.
This connection ``explains'' the origin of the terms linear in $\psi_1$ and $\phi_1$
in the dilaton in \rf{sol5}, relating them to particular terms in the dilaton of the gauged WZW model before the limit.
Note that to establish this connection we needed to start with the
deformation of the T-dual $AdS_n \times S^n$ background. For $n=5$ the ``reversal'' of the T-duality
is still not possible even for $\vk=i$ because of the linear
terms in $\psi_1$ and $\phi_1$ in the dilaton.

\subsection{pp-wave \texpdf{$\vk\to i$}{kappa -> i} limit and connection with Pohlmeyer reduction}\label{3.2}

It was shown in \ci{hrt} that the $\varkappa \to i$ limit can be made
more non-trivial by combining it with a particular rescaling of the directions $x^\pm = t \mp \vp$
as follows
\begin{equation}\la{kailim}
t = \epsilon \xxp + \frac{\xxm}{\epsilon} \ , \qquad
\varphi = \epsilon \xxp - \frac{\xxm}{\epsilon} \ , \qquad\qquad
\varkappa = i\sqrt{1 + \epsilon^2}\ , \qquad \epsilon \to 0\ .
\end{equation}
Then taking the $\epsilon \to 0$ limit in the solutions \eqref{sol2}, \eqref{sol3}, \eqref{sol5}
one finds
the following pp-wave backgrounds ($n=2,3,5$)\foot{If we
send $\epsilon \to i \epsilon$ and $x^\pm \to \pm i x^\pm$
we end up with the same solution with the opposite
sign for the coefficient of $(dx^+)^2$ and an imaginary flux.
As we will discuss, the light-cone gauge-fixing of \eqref{mo2} is
related to the Pohlmeyer-reduced theories for $AdS_n \times S^n$.
In an analogous way, the light-cone gauge-fixing of these backgrounds
after the analytic continuation, is related
to the Pohlmeyer-reduced theories for $dS_n \times H^n$.}
\ba\label{mo2}
&& ds_n^2 = -4d\xxp d\xxm + {\textstyle \frac12}(\cos 2\alpha - \cosh 2\beta)(d\xxm)^2 + ds_{n\perp}^2 \ , \\
&& F_n = dC_{n-1} =
d \CC_{n-2}
\W d\xxm \ , \qquad \qquad \qquad\
\Phi_n = \Phi_{n\perp} \ . \la{mm2}
\eea
Here the transverse metric $ds_{n\perp}^2$ and the dilaton $\Phi_{n\perp} $
are the same as in \eqref{1} (i.e. given by \rf{d2}--\rf{d5s}).
Note that since in the limit \rf{kailim} one has $t + \vp =2 \epsilon x^- \to 0$
the longitudinal part of the dilaton in \rf{1} is absent, i.e. the
dilaton here depends only on the transverse coordinates.
The RR potential
$C_{n-1}$
also depends only on the transverse coordinates $x_\perp= (\alpha, \beta, ...)$, where again we use \rf{rrr}.

Let us present the explicit form of the ``null'' RR backgrounds $F_n$ in \rf{mm2}.
In the $AdS_2 \times S^2$ case, which was discussed in \ci{ht}, from \rf{sol2} one finds\foot{This matches the
expression in \ci{ht} with a symmetric choice of the free constants there.}
\be \la{319}
C_1 = \sqrt 2 e^{-\P_0} ( \cos \a\, \sinh \b + \sin \a\, \cosh \b)  d\xxm
\ . \ee
In this case it was shown in \cite{hrt} that the
light-cone gauge-fixing ($\xxm = \mu \tau$) of the string in the background \eqref{mo2}, \eqref{d2}, \eqref{mm2}, \eqref{319}
yields the corresponding PR model, which is equivalent
\cite{gt} to the $\mathcal{N}=2$ supersymmetric sine-Gordon model.

In the $AdS_3 \times S^3$ case, the transverse metric and dilaton were given in \rf{d3}.
From \rf{sol3} we find
\be \la{320}
C_2 = e^{-\P_0} ( \cos^2 \a\, \sinh^2 \b\, d \psi_1 - \sin^2 \a\, \cosh^2 \b \, d \phi_1 )\wedge d\xxm
\ . \ee
T-dualizing in $\psi_1$ and $\phi_1$, or, alternatively, considering the analytic continuation
$x^\pm \to \pm i x^\pm$, $\alpha \to \alpha + \frac{\pi}{2}$, $\beta \to \beta + \frac{i\pi}{2}$, $\Phi_0 \to \Phi_0 + \frac{i\pi}{2}$,
we recover the ``null'' 3-form background supporting the corresponding pp-wave metric
\rf{mo2}, \rf{d3} that was found directly in \ci{hrt} (see also \ci{bak} for a similar 10d
solution supported by $F_5$).
The light-cone gauge-fixed form of the resulting string sigma model
is equivalent \cite{hrt} to the axial-gauged version of the
PR model for $AdS_3 \times S^3$ \ci{gt}. Therefore,
the light-cone gauge-fixing of this model before T-dualizing or analytically continuing
is equivalent to the vector-gauged version of the PR model.

As was explained above, in the $AdS_5 \times S^5$ case the transverse metric and dilaton
in \rf{mo2}, \rf{mm2} are the same as the special
limit of the metric-dilaton background for the $[SO(5)/SO(4)] \times [SO(1,4)/SO(4)$]
gauged WZW model.
The latter defines the bosonic
kinetic part of the PR model associated to the $AdS_5\times S^5$ superstring
\ci{gt,ms}.\foot{Note that, as for the
$n=2,3$ cases \cite{hrt,ht}, the $\vk \to i$ limit of the deformed $AdS$ part of the background is associated to the PR model
for the sphere and vice versa.}
Thus (in contrast to the $n=2$ and $3$ cases) in the $n=5$ case
the light-cone gauge-fixed action of the
superstring model corresponding to the background \rf{mo2}, \rf{mm2}
should be only a
{\it limit} of the full $AdS_5\times S^5$ PR model,
in agreement with the picture suggested in \ci{ht}.

In the $n=5$ case we find from \rf{sol5} that $F_5=dC_4$ with\foot{Note
that $C_4$ takes a simple form when written in terms of the diagonal-metric coordinates
used in \rf{101}, i.e.
$y^2= \sin^2 \z\, e^{2i\psi_1}$, $x^2=e^{2i\psi_1}$ and
$ {\rm y}^2= \sin^2 \xi\, e^{{\reo -} 2i\phi_1}$, ${\rm x}^2=e^{{\reo -} 2i\phi_1} $.}
{\begin{equation}\begin{split}\label{321}
C_4 = {\reo -\,} i e^{-\Phi_0}\Big[& \cos^4\a\,\sinh^4\b\,
d\psi_2 \W d(e^{2i\psi_1})\W d (\sin^2\xi\, e^{{\reo -}2i\phi_1})
\\ &
+ \sin^4\a \, \cosh^4 \b \, d( \sin^2 \z\, e^{2i\psi_1}) \W d\phi_2 \W d(e^{{\reo -}2i\phi_1})\Big] \wedge d\xxm\ .
\end{split}
\end{equation} }
As for the transverse metric and dilaton \eqref{mo2}, \eqref{mm2} the corresponding
5-form can be made real by analytically continuing $\psi_1 \to i \psi_1$, $\phi_1 \to i\phi_1$, along
with shifting $\Phi_0$ by $\frac{i\pi}{2}$.

To conclude, we have seen that in each of the \adt and \adst cases there is a
background for which the corresponding superstring theory taken in light-cone
gauge gives precisely the PR model for the original $AdS_n \times S^n$
superstring. The PR model was found by solving the conformal-gauge constraints
of the $AdS_n \times S^n$ superstring at the classical level using a non-local
change of variables and then reconstructing a new local action.\foot{This
connection gives a new perspective on the properties of the PR model like its
UV finiteness and the possibility of world-sheet supersymmetry. pp-waves with
curved transverse space in general do not preserve space-time supersymmetry and
thus a priori one should not expect explicit world-sheet supersymmetry for the
$n > 2$ cases, although there may be a hidden one (see \ci{hrt,bak} for related
discussions).}

At the same time, in the $n =5$ case the situation is different; we first need
to take a certain limit of the \adss PR model (generating 2+2 isometries in the
kinetic terms, cf. \rf{d5}, \rf{d5s}) in order to relate it to the light-cone
gauge-fixed superstring on the pp-wave background \rf{mo2}, \rf{mm2}, \rf{321}.
Originating directly from light-cone gauge-fixed pp-wave superstring, this
``limiting'' PR model should have some special features and deserves further
investigation.

\section{Summary and concluding remarks}

In this paper we have found a type IIB supergravity solution (with only the
metric, dilaton and 5-form being non-trivial) that can be interpreted as
one-parameter ($\vk$ or $\eta$ \rf{01}) deformation of the background obtained
from the maximally symmetric \adss background by applying T-duality in all 6
isometric directions. The latter ($\vk=0$) solution has imaginary RR flux,
which is a consequence of the formal T-duality being applied in time and this
feature persists for $\vk\not=0$. Another unusual property of the solution
\rf{sol5} is that for $\vk\not=0$ the dilaton has a linear term $\Phi_{5\, \rm
lin} =- 4\vk( t + \vp) {\,\reo -\,} 2\vk( \psi_1 - \phi_1)$ depending on a
linear combination of 4 out of 6 isometric directions of the metric.

Still, the metric and $e^{\Phi_5} F_5$, which enter the corresponding classical
Green-Schwarz superstring action, are invariant under shifts in the isometric directions
(i.e. these coordinates enter the Green-Schwarz action only through their derivatives) and
so one can formally T-dualize in them, as, e.g., in \ci{tds} -- assuming one
can first ignore the non-invariant linear piece in the ``quantum'' dilaton term
of the action. Remarkably, the resulting T-dual sigma model is equivalent (at
least to quadratic order in fermions) to the $\eta$-deformation \ci{dmv} of the
\adss superstring, i.e. the T-dual background has exactly the same metric,
$B$-field and combination of RR fluxes with the dilaton $e^{\hat{\Phi}_5}
\hat F_n$ ($n=1,3,5)$ as found in \ci{abf,abf2} directly from the action of the \emo
of \ci{dmv}. However, the presence of $\Phi_{5\, \rm lin}$ in the dilaton,
which depends on the what are now ``dual'' coordinates means that, in contrast
to the solution \rf{sol5} found here, the T-dual background of \ci{abf,abf2}
cannot be directly interpreted as (the non-dilaton part of) a standard type IIB
supergravity solution -- the full quantum T-dual sigma model including the
dilaton term appears to be defined on a ``doubled'' space.\foot{The usual
T-duality transformation at the level of type II supergravity maps solutions to
solutions but it applies only in the presence of an (abelian) isometry; upon
compactifying on an isometric direction the T-duality becomes equivalent to a
field redefinition. This logic does not apply in the case when the dilaton
depends on isometric directions of the metric, suggesting one should start with
a ``doubled'' string theory extension of type IIB supergravity, cf.
\ci{ddd,doub}.
Another possible idea for bypassing the complication of the linear 
non-isometric dilaton is to replace it in the action with the term
$e^{-2 \Phi_{5\, \rm lin}} \del_+ u \del_- v$, where $u$ and $v$ are two extra
coordinates; integrating out $u$ and $v$ produces the dilaton shift equal to
$\Phi_{5\, \rm lin}$. Redefining $u$ and $v$ by $e^{-\Phi_{5\, \rm lin}}$
the total action will depend only on derivatives of the isometric coordinates
so one will be able to T-dualize in them. However, 
one will not be able to easily  integrate out $u,v$  after doing the T-duality, i.e.  
the resulting background
will be a 12-dimensional one (with signature $-,-,+,\ldots,+$) and hence
its interpretation is unclear.}
The precise meaning of the relation between our solution
\rf{sol5}, the background of \ci{abf,abf2} and the \emo of \ci{dmv} thus
remains to be clarified.\foot{This (partial) T-duality relation between our
solution and the background of \ci{abf,abf2} appears to imply that the \emo of
\ci{dmv} should be one-loop UV finite at least in the bosonic sector. Indeed,
the Green-Schwarz superstring action corresponding to a type II supergravity background
should be Weyl invariant, and, in particular, UV finite. This should be true
for the Green-Schwarz model built on our solution. Since the formal T-dual of the latter
corresponds to the metric, $B$-field and $e^\Phi F_k$ background of
\ci{abf,abf2}, i.e. leads to the bosonic part and quadratic fermionic terms of
the \emo action, the conditions of scale invariance of the \emo in the metric
part, i.e. the generalized Einstein equations modulo reparametrizations
($R_{\m\n} + \sum (e^\Phi F_k)^2_{\m\n} + \ldots =\nabla_\m \xi_\nu + \nabla_\n
\xi_\m$), should be satisfied regardless of the dilaton issue.}

The RR flux vanishes at the special value of the deformation parameter $\vk =
i$, at which point our background factorizes into the product of the three
factors: a flat 2d ``longitudinal'' space $(t,\vp)$ with linear imaginary
dilaton $-4i(t + \vp)$ in \rf{1} and two 4d ``transverse'' metric-dilaton
backgrounds \rf{d5} and \rf{d5s}. Each of these factors represents a conformal
bosonic sigma model and they also solve the 10d supergravity equations, with
central charge shifts cancelling between the two 4d models. We have shown that
the two 4d factors can be interpreted as special limits \ci{ht} of the
backgrounds corresponding to the $SO(5)/SO(4)$ and $SO(1,4)/SO(4)$ gauged WZW
models respectively. This relates the linear terms in the two dilatons to
``blowing up'' certain angular factors in the standard gauged WZW dilatons.

We have also observed that these ``transverse'' backgrounds may be viewed as
defining the kinetic term of a limit \ci{ht} of the Pohlmeyer-reduced model
associated with the \adss superstring. We still cannot (due to the linear
dilaton term) T-dualize back to relate this to the $\vk=i$ limit of the \emo
directly, however, we can formally do so at the level of the Green-Schwarz
action. As shown in \cite{abf} the \emo is associated to the vertex
(particle-like) q-deformed S-matrix of \ci{bk}, in particular in the $\vk \to
i$ limit it should be associated to the limit of that S-matrix investigated in
\cite{prvs}. The full PR model should, however, be associated with the soliton
S-matrix. Therefore, this limit may be implementing the ``soliton-like picture
to particle-like picture'' transformation \ci{hhm}. It would be interesting to
clarify if this is indeed the case.

Finally, let us note that according to the discussion in \ci{ht} we should
expect a relation between the solution found here and the one constructed in
\ci{dst} that supports the metric of the $\l$-model, i.e. a one-parameter
deformation of the non-abelian dual of \adss superstring. Such a relation is
known to hold in the \adt case \ci{ht}. The background of \ci{dst} also
contains only the metric, dilaton and 5-form flux and therefore, may indeed
reduce to \rf{sol5} after taking a limit and doing an analytic continuation.
The two metrics are related by such a procedure \ci{ht} but the precise
correspondence between the dilatons and 5-forms remains to be checked.

\section*{Acknowledgments}
We would like to thank G. Arutyunov, R. Borsato, S. Frolov, T. Hollowood, J. L.
Miramontes, O. Lunin, R. Roiban, K. Sfetsos, D. Thompson and S. van Tongeren
for discussions of related issues. We also thank G. Arutyunov, R. Borsato and
S. Frolov for comments on the draft. The work of BH was funded by the DFG
through the Emmy Noether Program ``Gauge Fields from Strings'' and SFB 647
``Space - Time - Matter. Analytic and Geometric Structures.'' The work of AAT
was supported by the ERC Advanced grant No.290456, STFC Consolidated grant
ST/J0003533/1 and RNF grant 14-42-00047.

\appendix

\section{6-fold T-dual of the \texpdf{$AdS_5 \times S^5$}{AdS5 x S5} \texpdf{$\eta$}{eta}-model background}\label{appa}
\def\theequation{A.\arabic{equation}}
\setcounter{equation}{0}

In this Appendix we give the details of the formal T-duality transformations of
the \emo metric, $B$-field and RR fluxes of \cite{abf2} that can be
used to establish a relation to our supergravity solution \eqref{sol5}. The
``background'' fields found in \cite{abf2} by assuming that the $\eta$-model
action of \ci{dmv} can be interpreted as a classical Green-Schwarz sigma model
in a particular type IIB supergravity background are given by\foot{It 
is interesting to note that reversing the signs of $\opsi_2$ and $\ophi_2$
reverses the sign of all the RR $k$-form strengths, while reversing the
signs of
all the isometric directions reverses the signs of the $B$-field along with
the RR 1- and 5-forms. Combining the two transformations reverses the signs of
the $B$-field and the RR 3-form. These all correspond to well-known
$\mathbb{Z}_2$ symmetries of Type IIB theory.}
\small
{\allowdisplaybreaks
\begin{align}\nonumber
d\hat{s}_5^2 & = -\frac{1+\rho^2}{1-\varkappa^2\rho^2} d\ot^2 + \frac{d\rho^2}{(1-\varkappa^2\rho^2)(1+\rho^2)} + \frac{\rho^2\cos^2\zeta}{1+\varkappa^2 \rho^4\sin^2\zeta}d\opsi_1^2 + \frac{d\zeta^2}{1+\varkappa^2 \rho^4 \sin^2 \zeta} +\rho^2 \sin^2\zeta d\opsi_2^2
\\\nonumber & \hspace{12pt} +\frac{1-r^2}{1+\varkappa^2r^2}d\ovarphi^2 + \frac{dr^2}{(1+\varkappa^2r^2)(1-r^2)} + \frac{r^2 \cos^2\xi}{1+\varkappa^2 r^4 \sin^2 \xi} d\ophi_1^2 + \frac{d\xi^2}{1+\varkappa^2 r^4 \sin^2 \xi} + r^2 \sin^2\xi d\ophi_2^2\ ,
\\\nonumber
\hat{B} & = \frac{\varkappa \rho^4 \sin \zeta\cos\zeta}{1+\varkappa^2 \rho^4 \sin^2\zeta} d\opsi_1 \W d\zeta - \frac{\varkappa r^4 \sin\xi\cos\xi}{1+\varkappa^2 r^4 \sin^2\xi} d\ophi_1 \W d \xi \ , 
\\\nonumber
e^{\hat{\Phi}} \hat{F}^{(1)} & = \vk^2 \X\,
\Big[\rho^4 \sin^2\zeta \,d \hat\psi_2 - r^4 \sin^2 \xi\, d\hat \phi_2\Big] \ ,
\\\nonumber
e^{\hat{\Phi}} \hat{F}^{(3)} & =\vk\, \X\,
\Big[\frac{\rho^3\sin^2\zeta}{1-\vk^2\r^2} d\ot\wedge d\opsi_2 \wedge d\r + \frac{r^3 \sin^2\xi}{1+\vk^2 r^2} d\ovarphi\wedge d\ophi_2\wedge dr
\\\nonumber
& \hspace{30pt}
+\frac{\r^4\sin\z\cos\z}{1+\vk^2\r^4 \sin^2\z}d\opsi_2\W d\opsi_1 \W d\zeta
+\frac{r^4\sin\xi\cos\xi}{1+\vk^2 r^4 \sin^2\xi}d\ophi_2\W d\ophi_1 \W d\xi
\\\nonumber
& \hspace{30pt} + \frac{\vk^2\r r^4\sin^2\xi}{1-\vk^2\r^2}d\ot \W d\r \W d\ophi_2 - \frac{\vk^2 \r^4 r \sin^2\z}{1+\vk^2 r^2} d\opsi_2\W d \ovarphi \W d r
\\\nonumber
& \hspace{30pt}+\frac{\vk^2\r^4 r^4\sin\z\cos\z\sin^2\xi}{1+\vk^2 \rho^4 \sin^2 \z} d\opsi_1 \W d\z \W d \ophi_2
+\frac{\vk^2\r^4 r^4\sin^2\z\sin\xi\cos\xi}{1+\vk^2 r^4 \sin^2 \xi} d\opsi_2 \W d \ophi_1\W d \xi \Big]\ ,
\\\nonumber
e^{\hat{\Phi}} \hat{F}^{(5)} & = \X\,
\Big[
\frac{\r^3\sin\z\cos\z}{(1-\vk^2\r^2)(1+\vk^2\r^4\sin^2\z)}d\ot \W d\opsi_2 \W d\opsi_1 \W d\z \W d \r
\\\nonumber
&\hspace{30pt}
- \frac{r^3\sin\xi\cos\xi}{(1+\vk^2 r^2)(1+\vk^2 r^4 \sin^2 \xi)} d \ovarphi \W d \ophi_2 \W d \ophi_1 \W d\xi \W dr
\\\nonumber
&\hspace{30pt}
-\frac{\vk^2\r r}{(1-\vk^2\r^2)(1+\vk^2 r^2)} (\r^2 \sin^2 \z\, d\ot \W d \opsi_2 \W d\r \W d\ovarphi \W dr
+ r^2 \sin^2 \xi \, d\ot \W d\r \W d\ovarphi \W d\ophi_2 \W dr)
\\\nonumber
&\hspace{30pt}
+ \frac{\vk^2\r^4 r^4\sin\z\cos\z \sin\xi \cos\xi}{(1+\vk^2\r^4\sin^2\z)(1+\vk^2 r^4 \sin^2 \xi)} 
(d\opsi_2 \W d\opsi_1 \W d\z \W d\ophi_1 \W d\xi - 
d\opsi_1 \W d\z \W d\ophi_2 \W d\ophi_1 \W d\xi) 
\\\nonumber
&\hspace{30pt}
+ \frac{\vk^2\r r^4 \sin\xi\cos\xi}{(1-\vk^2\r^2)(1+\vk^2 r^4 \sin^2\xi)}(\r^2\sin^2\z\,d\ot \W d\opsi_2 \W d\r \W d\ophi_1 \W d\xi
- d\ot \W d\r \W d\ophi_2 \W d\ophi_1 \W d\xi)
\\\nonumber
&\hspace{30pt}
-\frac{\vk^2 \r^4 r\sin\z\cos\z}{(1+\vk^2 r^2)(1+\vk^2 \r^4 \sin^2 \z)} (r^2\sin^2\xi\, d\opsi_1 \W d\z \W d \ovarphi \W d\ophi_2 \W dr
+ d\opsi_2 \W d\opsi_1 \W d\z \W d \ovarphi \W dr)
\\\nonumber
& \hspace{30pt} -\frac{\vk^4 \r^5 r^4 \sin\z\cos\z\sin^2\xi}{(1-\vk^2\r^2)(1+\vk^2\r^4\sin^2\z)} d\ot \W d\opsi_1 \W d\z \W d\r\W d\ophi_2
\\\nonumber
& \hspace{30pt}
- \frac{\vk^4 \r^4 r^5 \sin^2\z \sin\xi\cos\xi}{(1+\vk^2 r^2)(1+\vk^2 r^4 \sin^2 \xi)} d\opsi_2 \W d\ovarphi \W d\ophi_1 \W d\xi \W dr\Big] \ , \\
& \X \equiv \frac{4 \sqrt{1+\vk^2}}{\sqrt{1-\vk^2\r^2}\sqrt{1+\vk^2\r^4\sin^2\z}\sqrt{1+\vk^2r^2}\sqrt{1+\vk^2r^4\sin^2\xi}} \ . \label{abfbackground}
\end{align}}\normalsize
Here $\hat \Phi $ is some a priori unknown dilaton and $\hat F^{(k)}$ are RR $k$-form strengths of type IIB theory.
The self-duality equation for the RR 5-form used in \cite{abf2} has the opposite sign 
to that used in \rf{seom5} in  section \ref{secsugra5}, that is\foot{Recall that  $\epsilon_{0123456789} = -1$ and that 
$m,n,\ldots = 0,1,2,3,4,5,6,7,8,9$ correspond to $ t,\psi_2,\psi_1,\z,\r,\vp,\phi_2,\phi_1,\xi,r$.}
\begin{equation}
\hat{F}^{(5)}_{mnpqr} = \frac{1}{5!}\sqrt{-\hat g}\,\e_{mnpqrstuvw}\hat{F}^{(5)}{}^{stuvw} \ .
\end{equation}
We also introduce the usual 7- and 9-forms (defined in terms of the dual RR 3- and 1-forms)
\begin{equation}
\hat{F}^{(7)}_{mnpqrst} = - \frac{1}{3!}\sqrt{-\hat g}\,\e_{mnpqrstuvw}\hat{F}^{(3)}{}^{uvw} \ , \qquad
\hat{F}^{(9)}_{mnpqrstuv} = \sqrt{-\hat g}\,\e_{mnpqrstuvw}\hat{F}^{(1)}{}^{w} \ .
\end{equation}


For the conventions of type IIB theory used in \cite{abf2}, the T-duality
transformation rules can be presented as follows \cite{hull,hass}
(here $\hat{y}$ stands for an isometric direction along which we dualize,
while $y$ is the dual coordinate)
\def \vp {{\vphantom{(-1)}}}
\begin{align}\nonumber
& g_{yy}^\vp = \hat{g}^{-1}_{\hat y\hat y} \ , \qquad g_{ym}^\vp = 
 \hat{g}_{\hat y\hat y}^{-1} \hat{B}_{\hat ym}^\vp\ , \qquad \qquad
B_{ym}^\vp = 
\hat{g}_{\hat y\hat y}^{-1} \hat{g}_{\hat ym}^\vp \ , \qquad
e^{\Phi} = \hat{g}_{\hat y\hat y}^{-1/2}e^{\hat\Phi} \ ,
\\\nonumber
&g_{mn}^\vp = \hat{g}_{mn}^\vp - \hat{g}_{\hat y\hat y}^{-1} (\hat g_{\hat ym}^\vp\hat g_{\hat yn}^\vp - \hat B_{\hat ym}^\vp\hat{B}_{\hat yn}^\vp) \ , \qquad
B_{mn}^\vp = \hat{B}_{mn}^\vp - \hat{g}_{yy}^{-1} (\hat g_{\hat ym}^\vp\hat B_{\hat yn}^\vp - \hat B_{\hat ym}^\vp\hat{g}_{\hat yn}^\vp) \ ,
\\\nonumber
&F^{(k)}_{ym_2\ldots m_k} =
- \hat{F}^{(k-1)}_{m_2\ldots m_k} + (k-1)\hat{g}_{\hat y\hat y}^{-1} \hat{g}_{\hat y[m_2}^\vp\hat{F}^{(k-1)}_{\hat ym_3\ldots m_k]} \ ,
\\\label{rules}
& F^{(k)}_{m_1 m_2\ldots m_k} = - \hat F^{(k+1)}_{\hat ym_1\ldots m_k} 
+ k \hat B_{\hat y[m_1}^{\vp}F^{(k)}_{ym_2\ldots m_k]} \ ,
\end{align} 
while the transformation rules for the RR potentials are similar to those of the RR fluxes.
Note that these relations imply that the vielbein components of $e^{\Phi} F$
get mapped into vielbein components of $e^{\hat \Phi}\hat F$, in agreement
with how T-duality acts on the quadratic fermionic term in  the Green-Schwarz action \cite{tds}. 

To perform the T-duality in time we first analytically continue $\hat t =i \hat t_E$,
then use the standard T-duality rules, and subsequently analytically continue
back $t_E = - i t$ \cite{Hull}. This introduces the required factor of $i$ in the RR fluxes.

We can formally apply these rules to the background fields in
\eqref{abfbackground} with the transformation of the combination of RR flux and
dilaton, $e^{\hat\Phi} \hat F^{(n)}$, found by combining the transformations of
the individual factors/components.
To compensate for the different choice of
self-duality equation we first reverse the sign of $\hat \psi_2$ in
\eqref{abfbackground}, 
after which, applying the T-duality rules \eqref{rules}, we recover our solution
\eqref{sol5}.

To summarize, even though the background fields in \eqref{abfbackground} cannot be extended
to a solution of type IIB supergravity and the fluxes are inconsistent with the Bianchi
identities \ci{abf2}, those found after applying the T-duality rules \eqref{rules} can be and are.
This is not in contradiction with the standard logic that T-duality maps from one
background solving the supergravity equations of motion and Bianchi identities to
another, as this assumes that the directions in which one dualizes are isometries of the 
full background, and not just the combinations of fields appearing in the classical
Green-Schwarz action.

\np




\begin{thebibliography}{30}
\parskip=0.2 pt

\bibitem{dmv}
F.~Delduc, M.~Magro and B.~Vicedo,
``An integrable deformation of the $AdS_5 \times S^5$ superstring action,''
Phys.\ Rev.\ Lett.\ {\bf 112}, no. 5, 051601 (2014)
[\arxivlink{1309.5850}].
``Derivation of the action and symmetries of the $q$-deformed $AdS_5 \times S^5$ superstring,''
JHEP {\bf 1410} (2014) 132
[\arxivlink{1406.6286}].

\bibitem{klim}
C.~Klimcik,
``Yang-Baxter sigma models and dS/AdS T duality,''
JHEP {\bf 0212}, 051 (2002)
[\arxivlink{hep-th/0210095}].
``On integrability of the Yang-Baxter sigma-model,''
J.\ Math.\ Phys.\  {\bf 50}, 043508 (2009)
[\arxivlink{0802.3518}].

\bibitem{abf}
G.~Arutyunov, R.~Borsato and S.~Frolov,
``S-matrix for strings on $\eta$-deformed $AdS_5 \times S^5$,''
JHEP {\bf 1404} (2014) 002
[\arxivlink{1312.3542}].

\bibitem{hrt}
B.~Hoare, R.~Roiban and A.~A.~Tseytlin,
``On deformations of $AdS_n \times S^n$ supercosets,''
JHEP {\bf 1406}, 002 (2014)
[\arxivlink{1403.5517}].

\bibitem{lrt}
O.~Lunin, R.~Roiban and A.~A.~Tseytlin,
``Supergravity backgrounds for deformations of $AdS_n \times S^n$ supercoset string models,''
Nucl.\ Phys.\ B {\bf 891}, 106 (2015)
[\arxivlink{1411.1066}].

\bibitem{abf2}
G.~Arutyunov, R.~Borsato and S.~Frolov,
``Puzzles of eta-deformed $AdS_5 \times S^5$,''
[\arxivlink{1507.04239}].

\bibitem{hms}
T.~J.~Hollowood, J.~L.~Miramontes and D.~M.~Schmidtt,
``An Integrable Deformation of the $AdS_5 \times S^5$ Superstring,''
J.\ Phys.\ A {\bf 47} (2014) 49, 495402
[\arxivlink{1409.1538}].
``Integrable Deformations of Strings on Symmetric Spaces,''
JHEP {\bf 1411} (2014) 009
[\arxivlink{1407.2840}].

\bibitem{sfet}
K.~Sfetsos,
``Integrable interpolations: From exact CFTs to non-Abelian T-duals,''
Nucl.\ Phys.\ B {\bf 880} (2014) 225
[\arxivlink{1312.4560}].

\bi{dmv2}
B.~Vicedo,
``Deformed integrable $\sigma$-models, classical $R$-matrices and classical exchange algebra on Drinfel'd doubles,''
[\arxivlink{1504.06303}].

\bibitem{ht}
B.~Hoare and A.~A.~Tseytlin,
``On integrable deformations of superstring sigma models related to $AdS_n \times S^n$ supercosets,''
Nucl.\ Phys.\ B {\bf 897} (2015) 448
[\arxivlink{1504.07213}].

\bibitem{Sfetsos:2014cea}
K.~Sfetsos and D.~C.~Thompson,
``Spacetimes for $\lambda$-deformations,''
JHEP {\bf 1412}, 164 (2014)
[\arxivlink{1410.1886}].

\bibitem{dst}
S.~Demulder, K.~Sfetsos and D.~C.~Thompson,
``Integrable $\lambda$-deformations: Squashing Coset CFTs and $AdS_5 \times S^5$.''
[\arxivlink{1504.02781}].

\bi{hull}
E.~Bergshoeff, C.~M.~Hull and T.~Ortin,
``Duality in the type II superstring effective action,''
Nucl.\ Phys.\ B {\bf 451}, 547 (1995)
[\arxivlink{hep-th/9504081}].

\bi{tds}
M.~Cvetic, H.~Lu, C.~N.~Pope and K.~S.~Stelle,
``T duality in the Green-Schwarz formalism, and the massless / massive IIA duality map,''
Nucl.\ Phys.\ B {\bf 573}, 149 (2000)
[\arxivlink{hep-th/9907202}].
B.~Kulik and R.~Roiban,
``T duality of the Green-Schwarz superstring,''
JHEP {\bf 0209}, 007 (2002)
[\arxivlink{hep-th/0012010}].
L.~F.~Alday, G.~Arutyunov and S.~Frolov,
``Green-Schwarz strings in TsT-transformed backgrounds,''
JHEP {\bf 0606}, 018 (2006)
[\arxivlink{hep-th/0512253}].

\bi{hass}
S.~F.~Hassan,
``T duality, space-time spinors and RR fields in curved backgrounds,''
Nucl.\ Phys.\ B {\bf 568}, 145 (2000)
[\arxivlink{hep-th/9907152}].
``Supersymmetry and the systematics of T duality rotations in type II superstring theories,''
Nucl.\ Phys.\ Proc.\ Suppl.\ {\bf 102}, 77 (2001)
[\arxivlink{hep-th/0103149}].

\bi{ddd}
A.~A.~Tseytlin,
``Duality symmetric closed string theory and interacting chiral scalars,''
Nucl.\ Phys.\ B {\bf 350}, 395 (1991).
D.~S.~Berman and D.~C.~Thompson,
``Duality Symmetric String and M-Theory,''
Phys.\ Rept.\ {\bf 566}, 1 (2014)
[\arxivlink{1306.2643}].

\bi{doub}
W.~Siegel,
``Superspace duality in low-energy superstrings,''
Phys.\ Rev.\ D {\bf 48}, 2826 (1993)
[\arxivlink{hep-th/9305073}].
C.~Hull and B.~Zwiebach,
``Double Field Theory,''
JHEP {\bf 0909}, 099 (2009)
[\arxivlink{0904.4664}].
G.~Aldazabal, D.~Marques and C.~Nunez,
``Double Field Theory: A Pedagogical Review,''
Class.\ Quant.\ Grav.\ {\bf 30}, 163001 (2013)
[\arxivlink{1305.1907}].
O.~Hohm, D.~Lust and B.~Zwiebach,
``The Spacetime of Double Field Theory: Review, Remarks, and Outlook,''
Fortsch.\ Phys.\ {\bf 61}, 926 (2013)
[\arxivlink{1309.2977}].

\bibitem{nun}
D.~Geissbuhler, D.~Marques, C.~Nunez and V.~Penas,
``Exploring Double Field Theory,''
JHEP {\bf 1306}, 101 (2013)
[\arxivlink{1304.1472}].

\bibitem{Hull}
C.~M.~Hull,
``Timelike T duality, de Sitter space, large N gauge theories and topological field theory,''
JHEP {\bf 9807}, 021 (1998)
[\arxivlink{hep-th/9806146}].

\bi{sor}
D.~Sorokin, A.~Tseytlin, L.~Wulff and K.~Zarembo,
``Superstrings in $AdS_2 \times S^2 \times T^6$,''
J.\ Phys.\ A {\bf 44}, 275401 (2011)
[\arxivlink{1104.1793}].

\bibitem{mirror0}
G.~Arutyunov and S.~J.~van Tongeren,
``$AdS_5 \times S^5$ mirror model as a string sigma model,''
Phys.\ Rev.\ Lett.\ {\bf 113} (2014) 261605
[\arxivlink{1406.2304}].
JHEP {\bf 1505} (2015) 027
[\arxivlink{1412.5137}].

\bi{gt}
M.~Grigoriev and A.~A.~Tseytlin,
``Pohlmeyer reduction of $AdS_5 \times S^5$ superstring sigma model,''
Nucl.\ Phys.\ B {\bf 800}, 450 (2008)
[\arxivlink{0711.0155}].
``On reduced models for superstrings on $AdS_n \times S^n$,''
Int.\ J.\ Mod.\ Phys.\ A {\bf 23} (2008) 2107
[\arxivlink{0806.2623}].

\bi{rabi}
K.~Bardakci, M.~J.~Crescimanno and E.~Rabinovici,
``Parafermions From Coset Models,''
Nucl.\ Phys.\ B {\bf 344}, 344 (1990).
E.~Witten,
``On string theory and black holes,''
Phys.\ Rev.\ D {\bf 44}, 314 (1991).

\bi{bs}
I.~Bars and K.~Sfetsos,
``A Superstring theory in four curved space-time dimensions,''
Phys.\ Lett.\ B {\bf 277}, 269 (1992)
[\arxivlink{hep-th/9111040}].
``Conformally exact metric and dilaton in string theory on curved space-time,''
Phys.\ Rev.\ D {\bf 46}, 4510 (1992)
[\arxivlink{hep-th/9206006}].

\bibitem{rt}
R.~Roiban and A.~A.~Tseytlin,
``UV finiteness of Pohlmeyer-reduced form of the $AdS_5 \times S^5$ superstring theory,''
JHEP {\bf 0904} (2009) 078
[\arxivlink{0902.2489}].

\bi{fl}
E.~S.~Fradkin and V.~Y.~Linetsky,
``On space-time interpretation of the coset models in D $<$ 26 critical string theory,''
Phys.\ Lett.\ B {\bf 277}, 73 (1992).

\bi{bak}
I.~Bakas and J.~Sonnenschein,
``On Integrable models from pp wave string backgrounds,''
JHEP {\bf 0212}, 049 (2002)
[\arxivlink{hep-th/0211257]}.

\bi{ms}
A.~Mikhailov and S.~Schafer-Nameki,
``Sine-Gordon-like action for the Superstring in $AdS_5 \times S^5$,''
JHEP {\bf 0805}, 075 (2008)
[\arxivlink{0711.0195]}.

\bi{bk}
N.~Beisert and P.~Koroteev,
``Quantum Deformations of the One-Dimensional Hubbard Model,''
J.\ Phys.\ A {\bf 41} (2008) 255204
[\arxivlink{0802.0777}].
B.~Hoare, T.~J.~Hollowood and J.~L.~Miramontes,
``q-Deformation of the $AdS_5 \times S^5$ Superstring S-matrix and its Relativistic Limit,''
JHEP {\bf 1203} (2012) 015
[\arxivlink{1112.4485}].

\bi{prvs}
B.~Hoare and A.~A.~Tseytlin,
``Towards the quantum S-matrix of the Pohlmeyer reduced version of $AdS_5 \times S^5$ superstring theory,''
Nucl.\ Phys.\ B {\bf 851} (2011) 161
[\arxivlink{1104.2423}].
B.~Hoare, T.~J.~Hollowood and J.~L.~Miramontes,
``A Relativistic Relative of the Magnon S-Matrix,''
JHEP {\bf 1111} (2011) 048
[\arxivlink{1107.0628}].

\bi{hhm}
B.~Hoare, T.~J.~Hollowood and J.~L.~Miramontes,
``Restoring Unitarity in the q-Deformed World-Sheet S-Matrix,''
JHEP {\bf 1310} (2013) 050
[\arxivlink{1303.1447}].

\end{thebibliography}
\end{document}